\documentclass[aps,pra,twocolumn,showpacs,amsmath,amssymb,preprintnumbers,superscriptaddress,10pt]{revtex4-1}

\usepackage{graphicx}
\graphicspath{{figures/}}

\usepackage{amsmath}
\usepackage{lipsum, color}

\usepackage{comment}
\usepackage{amsmath, amssymb}
\usepackage{bm}% bold math
\usepackage{physics}
\usepackage{graphicx}% include figure files
\usepackage{SIunits}
\usepackage{hyphenat}
\usepackage[bookmarks=false] {hyperref}
\usepackage{color}
\usepackage[capitalise]{cleveref}
\crefname{section}{Sec.}{Secs.}% APS style uses abbreviations
\Crefname{section}{Section}{Sections}

\usepackage{soul}
\usepackage[utf8]{inputenc}
\usepackage{booktabs}
\usepackage{tabularx}
\usepackage{ltablex}

\definecolor{ss_color}{rgb}{1,0,0}

\definecolor{rf_color}{rgb}{0,0,1}

\begin{abstract}
The ability of an eavesdropper to compromise the security of a quantum communication system by changing the angle of the incoming light is well-known. Randomizing the role of the detectors has been proposed to be an efficient countermeasure to this type of attack. Here we show that the proposed countermeasure can be bypassed if the attack is generalized by including more attack variables. Using the experimental data from existing literature, we show how randomization effectively prevents the initial attack but fails to do so when Eve generalizes her attack strategy. Our result and methodology could be used to security-certify a free-space quantum communication receiver against all types of detector-efficiency-mismatch type attacks.

\end{abstract}

\begin{document}
	
\title{A generalized efficiency mismatch attack to bypass detection-scrambling countermeasure}
\date{7 January, 2021}

\author{M A Ruhul Fatin}
\email{ruhulfatin103@gmail.com}
\affiliation{Department of Electrical and Electronic Engineering, Bangladesh University of Engg and Tech., Dhaka, Bangladesh}
\affiliation{Department of Electronics, Carleton University, Ottawa, ON, K1S 5B6, Canada}

\author{Shihan~Sajeed}
\email{shihan.sajeed@gmail.com}
\affiliation{Institute for Quantum Computing, University of Waterloo, Waterloo, ON, N2L~3G1 Canada}
\affiliation{Department of Physics and Astronomy, University of Waterloo, Waterloo, ON, N2L~3G1 Canada}
\affiliation{Department of Electrical and Computer Engineering, University of Toronto, M5S~3G4, Canada}

\maketitle

\section{Introduction}
Recent trends in quantum technologies suggest a future of quantum computers (QC) having superior computational power \cite{wknight2017,dcastel}. Such computational power can efficiently solve hard mathematical problems that are the foundations of security for certain public-key cryptosystems. QCs thus pose a serious threat to our current cryptographic infrastructure. One possible solution can be post-quantum cryptography~\cite{stanthesis,mceliece1978public,hoffstein1998ntru} -- classical algorithms thought to be secure against quantum attacks  -- but there is no mathematical proof that these algorithms provide information theoretic security. Thus, in an effort to fight quantum with quantum the trend is towards quantum cryptography~\cite{bennett1984,gisin2002,scarani2009} -- more popularly known as quantum key distribution (QKD).

QKD \cite{gisin2002,scarani2009} uses the laws of quantum mechanics to generate a secret key between two distant parties Alice and Bob. This key then can be used for encryption using one-time-pad and guarantee secure communication. In theory, QKD provides mathematical proof of security by modeling the device behaviors and using the laws of quantum mechanics. However, in practice, devices often behave differently than the assumed model, leaving a gap between theory and practice that can be exploited by an eavesdropper. This gap can be anywhere in the system implementation such as measurement devices~\cite{lydersen2010a,gerhardt2011}, monitoring systems~\cite{sajeed2015}, assumption in the security proofs~\cite{sajeed2016}, leakage of information~\cite{jain2014,sajeed2017,pinheiro2018}, change of characteristics~\cite{makarov2016,bugge2014}, imperfect sources~\cite{bennett1992b,xu2015a}, imperfect detector characteristics~\cite{makarov2006,sajeed2015a,rau2015,chaiwongkhot2019} etc. It is essential for QKD security to explore and identify these gaps and characterize them in order to assess the threat. In this work we analyze one such gap -- detector-efficiency mismatch~\cite{makarov2006,qi2007,sajeed2015a,chaiwongkhot2019}-- and analyze its effects.

A fundamental assumption in QKD security proofs is that the measurement outcomes should be independent of the measurement bases and Eve should not have any control over them. In ideal QKD, it is impossible for Eve to control the measurement outcomes without introducing errors called quantum bit error rate (QBER). However, in practice, there might be implementation vulnerabilities that allow Eve to have this control. For example, if there is a sensitivity mismatch among the detectors for a certain degree of freedom of the incoming photons, Eve can modify that degree of freedom so that one detector becomes more sensitive compared to another~\cite{qi2007,makarov2008,sajeed2015a,rau2015}. This can happen in the time degree of freedom: implementation vulnerability may make one detector more sensitive in a particular time window than the others. In this case,  Eve can shift the arrival time of certain pulses to coincide with that window. Thus, detection events occurring in that particular time-window have a higher chance of occurring in the sensitive detector and a bias is achieved. Similarly, if the detector sensitivity varies with spatial-mode of the incoming light~\cite{makarov2006}, Eve can send light at certain angle ($\phi$,$\theta$) to create a bias among the detector sensitivity and achieves a control. The demonstration of exploiting such spatial-mode-sensitivity-mismatch was shown in ~\cite{sajeed2015a,rau2015}. 

%The sensitivity mismatch is shown in \cref{detector} which shows a total mismatch i.e., one detector is completely insensitive while the other remains sensitive. It should be noted that this picture gives a very simplistic view and in practice, the mismatch is a partial mismatch, i.e., one detector is more sensitive than the other.

% Efficiency mismatch in photodetectors\cite{makarov2006, makarov2008} have been considered a source of imperfection that can be manipulated to attack QKD protocols and have been successful in several experimental models. Attack models like time shift attack\cite{qi2007} and spatial mode efficiency mismatch attack\cite{sajeed2015a} are based on this imperfection. In case of spatial mode mismatch, Eve can cut in line between Alice-Bob communication and tilt the beam in an angle $(\theta, \phi)$ such that there is a higher click probability in one of the photodetectors than others. This is possible because the optical alignments of the system can have inherent imperfections due to limited manufacturing precision. In this case, Eve can gain some amount of control of Bob's detection outcome and implement a faked-state attack\cite{lydersen2010a,makarov2005} without introducing any significant error.  

A countermeasure to this loophole, \emph{detector scrambling}, was proposed in Refs.~\cite{da2014safeguarding} that involves randomly changing the roles of the detectors to hash out any mismatch in the detection system and reduce efficiency mismatch. In this paper, we scrutinize the effectiveness of this countermeasure. In \cref{em}, we introduce and review some necessary details of the spatial-mode-efficiency-mismatch attack reported in ~\cite{sajeed2015a}. In \cref{dsc}, we simulate a detector scrambling countermeasure and show the countermeasure blocks the side-channel. Then in \cref{conc}, we show how the scrambling countermeasure can be bypassed by resorting to a more general attack strategy. We conclude in \cref{conc}.

\section{Review of detection efficiency mismatch}
\label{em}
\begin{figure}[htb]
a)
	\includegraphics[width = 0.7\columnwidth]{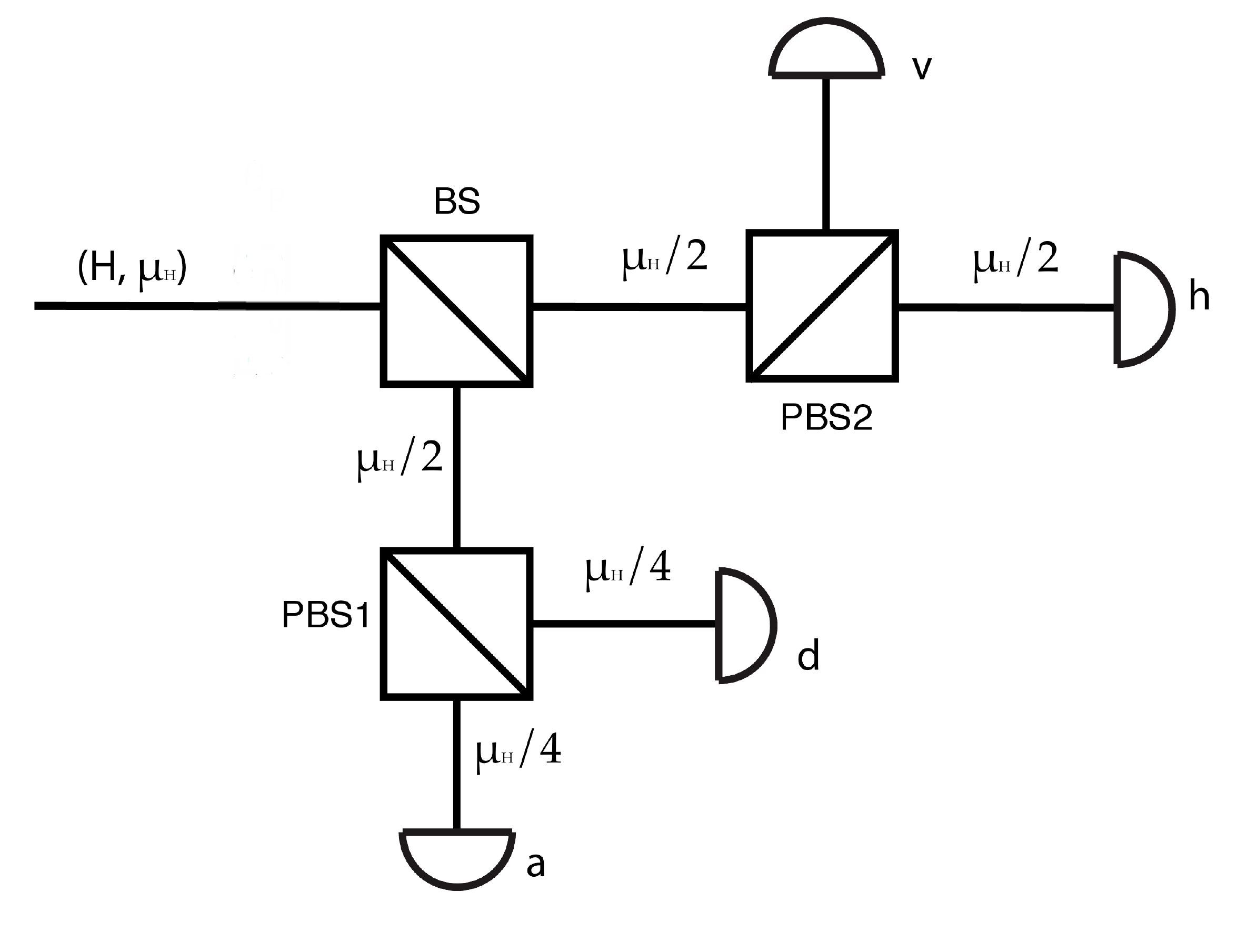}
\\b)
	\includegraphics[width = 0.7\columnwidth]{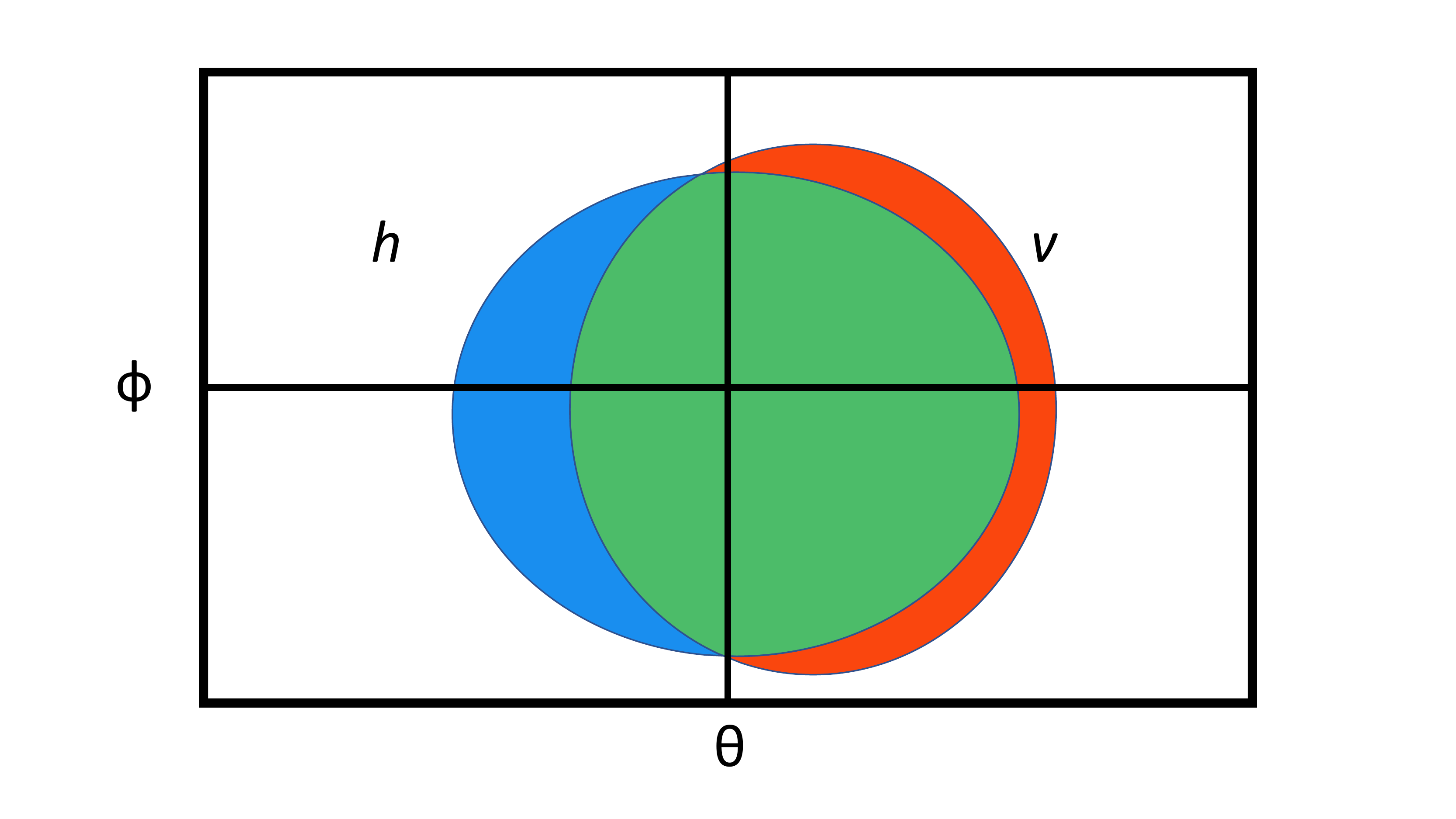}
	\caption{a) Schematic of a typical BB84 receiver setup with four photodetectors. The labels indicate the case when there is incoming $H$-polarized light.b) Spatial efficiency mismatch in detectors \emph{h} and \emph{v} when looked from the channel. Blue and Red regions indicate the efficiencies -- as a function of illumination angle ($\phi$, $\theta$) -- of the \emph{h} and \emph{v} detectors respectively where one detector is on while the other is off  . The green region indicates ranges of ($\phi$,$\theta$), where both detectors are equally sensitive. None of the detectors are sensitive in the white region. For a mismatch of this kind, an adversary can utilize this to get some information about the key.} 
	\label{setup1}
\end{figure}

We shall assume a polarization-encoded Bennett-Brassard (BB84) QKD scheme with passive basis-choice implementation as shown in \cref{setup1}a. The beam splitter (BS) is used for selecting the \emph{HV} or \emph{DA} bases and the polarization beam splitters (PBSs) followed by two detectors are used to measure the polarization in a basis. Detectors \emph{h}  and \emph{v} are used for measuring the incoming \emph{H} and \emph{V} polarized light while detectors \emph{d} and \emph{a} are used for measuring \emph{D} and \emph{A} polarized light respectively.

The efficiency-mismatch side-channel is explained with the help of \cref{setup1}b.  Here we show how the sensitivity of the $h$ and $v$ detectors varies in response to the angle of the incoming light. The circle on the left (right) shows the sensitive area of detector $h (v)$. Outside the circle the sensitivity is zero (in practical detectors, sensitivity does not go to zero so abruptly, but this simple assumption serves the purpose to explain the concept). In the overlapping (green) region, both the detectors are equally sensitive. However, if the light is sent towards the red (blue) region, detector $v (h)$ has a higher sensitivity than the $h (v)$ detector. Eve can stage a faked-stage attack to exploit this bias. 

The faked-state attack  considered in ref\cite{sajeed2015a} is based on the following assumptions. Eve is present outside Alice's lab. She intercepts and measures the signal going towards Bob. Then she reproduces another pulse with the same polarization as her measurement outcome but with different mean photon number,  and sends it towards Bob at an angle where the target detector has a higher sensitivity compared to others. More specifically, if Eve's measurement outcome is $j$, she reproduces $j$ polarized light with mean photon number $\mu_j$ and sends it at an angle where detector $j$ has a higher sensitivity than the other three detectors. This angle is referred to as the \emph{attack angle} for detector $j$. She uses a lossless channel to overcome the channel loss and maximize her target detection probabilities. The sifted key rate and QBER in Eve's presence become
\begin{equation}
\begin{aligned}
&R_e = \frac{1}{4} \sum_{j = H,V,D,A} R_e(j),\\
&\text{QBER}_e =\frac{1}{4 R_e} \sum_{j = H,V,D,A} E_j.
\end{aligned}
\label{r_qber}
\end{equation}

%\ssc{At this moment it is not clear where the efficiency mismatch term is hidden. that is why it is super important to introduce and properly explain the efficiency mismatch terms. for example what are $\eta_i(j)s$ }\ssc{You need to think from a reader's point of view and explain what is going on and why. You should talk more about the optimization, how Eve manages to steeal key. For that u need to introduce key rate and error rate without Eve also. Then say: Eve can match the two rates and still keep the errors to a minimum by manipulating four free parameters and angles. These are the key terms of the attack which you haven't talked about. After you have introduced the attack methodology sufficiently, only then you can move to the next section on how to prevent it.}\rf{I have introduced $R_{ab}$ which is the key rate without Eve in picture. I did not put the error rate without Eve because in our optimization program we have no use of error rate without Eve. We calculated the error rate with Eve in picture to find the overall QBER. We are keeping Rab=Re as our constraint equation and calculate the overall QBER with Eve in picture. Hope I have addressed all the issues here.}

In order to remain hidden, Eve's first target would be to match the sifted key rate $R_{e}$ to the expected key rate $R_{ab}$, i.e., $R_{ab} = R_{e}$. The next target would be to minimize  $QBER_e$ to maximize the amount of leaked information. Thus, the problem can be turned into an optimization problem with the goal of minimizing $QBER_e$ with the constraints $R_{ab} = R_{e}$. The parameters to optimize are the four mean photon numbers which Eve can manipulate to minimize the error.  A harder constraint can also be chosen. Instead of matching only total key rate, the key rate at each channel can also be matched.  Both of these optimizations were done in Ref.~\cite{sajeed2015a} and the result is reproduced in \cref{rand}.

\section{Detector Scrambling countermeasure }
\label{dsc}

In this section we discuss the general detector scrambling countermeasure outlined in \cite{da2014safeguarding} and investigate its effectiveness in preventing the attack. Let us assume that a half-wave plate (HWP) is placed in front of the BS in \cref{setup1}. By rotating the axis of the HWP Bob can rotate the incoming polarization by $\theta_{B} = 0~\degree, 45~\degree, 90~\degree$ and $135~\degree$. When $\theta_{B} = 0~\degree$, the detectors marked by \emph{h,v,d} and \emph{a} are used to detect incoming horizontal \emph{(H)}, vertical \emph{(V)}, diagonal \emph{(D)} and anti-diagonal \emph{(A)} polarized lights respectively. When $\theta_{B} = 90~\degree$ , the bases are unchanged but the roles of each detector is inverted, i.e, detector marked \emph{h} measures \emph{V} and vice versa. In case of $\theta_{B} = 45~\degree$, the roles of each basis is flipped and finally for $\theta_{B} = 135~\degree$ both the roles of each basis and each detector is flipped, i.e, a detector marked \emph{h} measures \emph{D} and \emph{A} when $\theta_{B} = 45~\degree$ and $\theta_{B} = 135~\degree$ respectively. Thus, by randomly changing the incoming polarization by a HWP, it is possible for Bob to scramble the roles of both his bases and detectors.  

%There has been reports of closing the back-door opened by the photodetectors having temporal efficiency mismatch\cite{Silva:2014aa} using this detector scrambling method. Using data from\cite{sajeed2015a} we are going to scrutinize this method in case of spatial-mode efficiency-mismatch of the photodetectors.

\begin{figure}[tb]
	\centering
	\includegraphics[width =0.8\columnwidth]{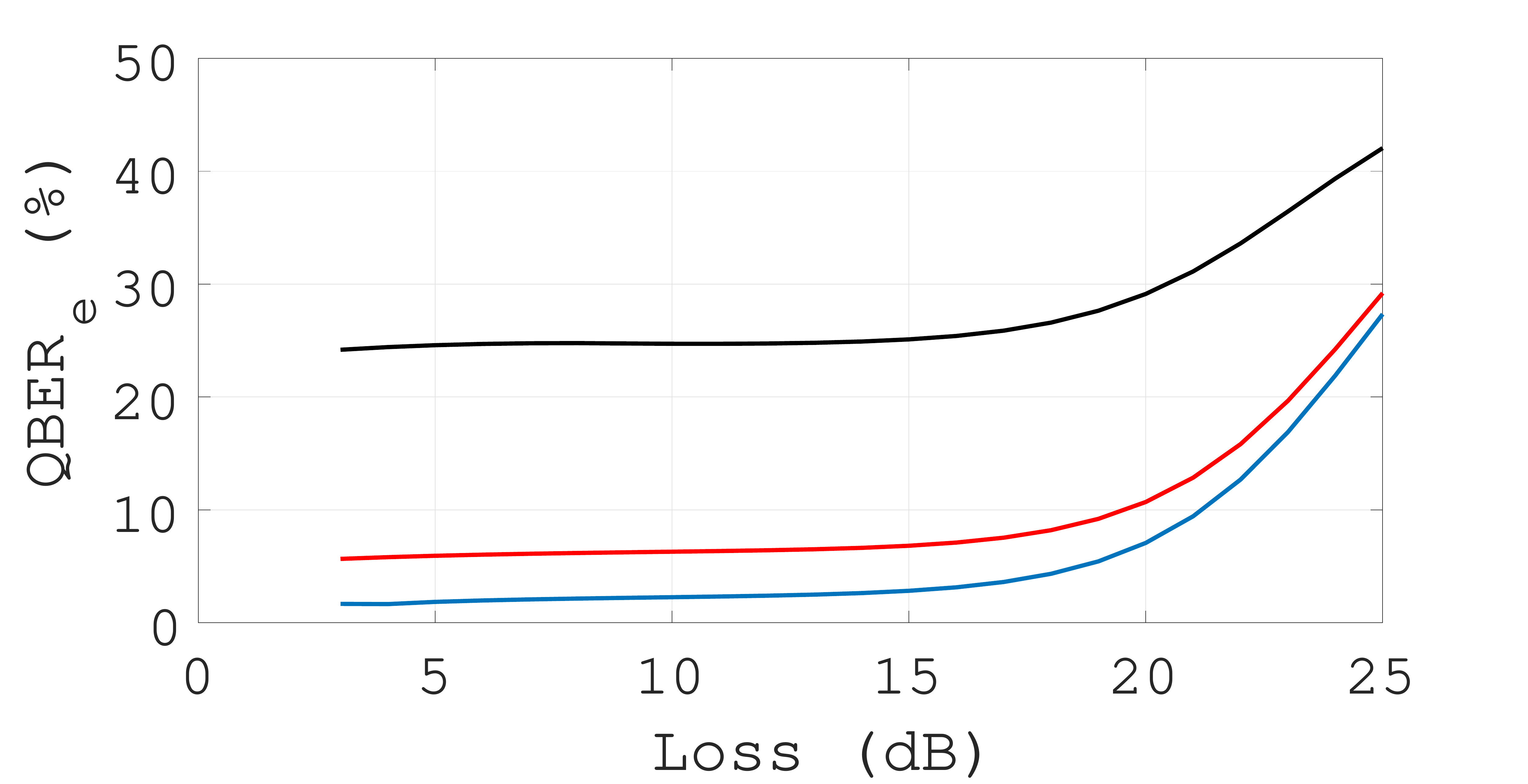}
	\caption{Simulated QBER vs line loss. The lower two solid curve (red and blue) indicates the results obtained in \cite{sajeed2015a}. The blue curve shows the optimized $QBER_e$ when Eve matches the Bob compares the total sifted key rate with the expected Alice-Bob sifted key rate $R_{ab}$. The red curve shows the optimized $QBER_e$ when Eve matches the rate for individual channels like $R_{ab} = R_{e}(j)$ where $j \in \{\mathrm{h,v,d,a}\}$. The upper solid curve shows the optimized QBER in presence of detector scrambling countermeasure. It can be seen that Bob can smoke out Eve's presence with this countermeasure.}
		\label{rand}
\end{figure}

In the following, we assume Bob scrambles his detectors with equal a-priori probability. The sifted key rate $R_e(j|\theta_B)$ and error rate $E_{j|\theta_B }$ in the presence of Eve given she sends $j$ polarized light -- towards attack angle $j$ with mean photon number $\mu_j$ --  and Bob applies $\theta_B$ rotation, can be derived similar to \cref{reh,pvhp,serror} as presented in \cref{scrambling}. Thus, the total sifted key rate $R_e^s$ and $QBER_e^s$  with Eve's attack and Bob applying scrambling countermeasure become (derived in \cref{scrambling}) :
\begin{equation}\label{equation_2}
\begin{split}
&R_e^s = \frac{1}{4}  \sum_{j = H,V,D,A} \frac{1}{4}\sum_{\theta= 0^{\circ},45^{\circ},90^{\circ},135^{\circ}} R_e(j | \theta)\\
&QBER_e^s = \frac{1}{4R_e} \sum_{j=H,V,D,A} \frac{1}{4}\sum_{\theta= 0^{\circ},45^{\circ},90^{\circ},135^{\circ}} E_{j | \theta},
\end{split}
\end{equation}
As discussed in \cref{em}, the terms $E_{j | \theta}$ and $R_e$ are dependent on mean photon number chosen by Eve. Thus, we perform similar optimization using the four mean photon numbers as the free parameters to minimize $QBER_e^s$ with the constraint $R_e^s= R_{ab}$. Our result is shown with the black curve in \cref{rand}. The presence of scrambling makes $QBER_e^s > 25\%$ and no successful key generation is possible. In the simulation, the efficiency of the detectors, the mismatch values, background counts and all other parameters are taken from~\cite{sajeed2015a}. This result highlights that as soon as Bob employs detector scrambling technique, Eve cannot manipulate the four mean photon numbers to achieve a QBER less than $25\%$ while satisfying the constraints of matching the rates. This shows the effectiveness of the scrambling countermeasure.

\section{Detector-scrambling-bypass strategy}
\label{dr}
\begin{figure}[b]
	\centering
	\includegraphics[width=0.8\columnwidth]{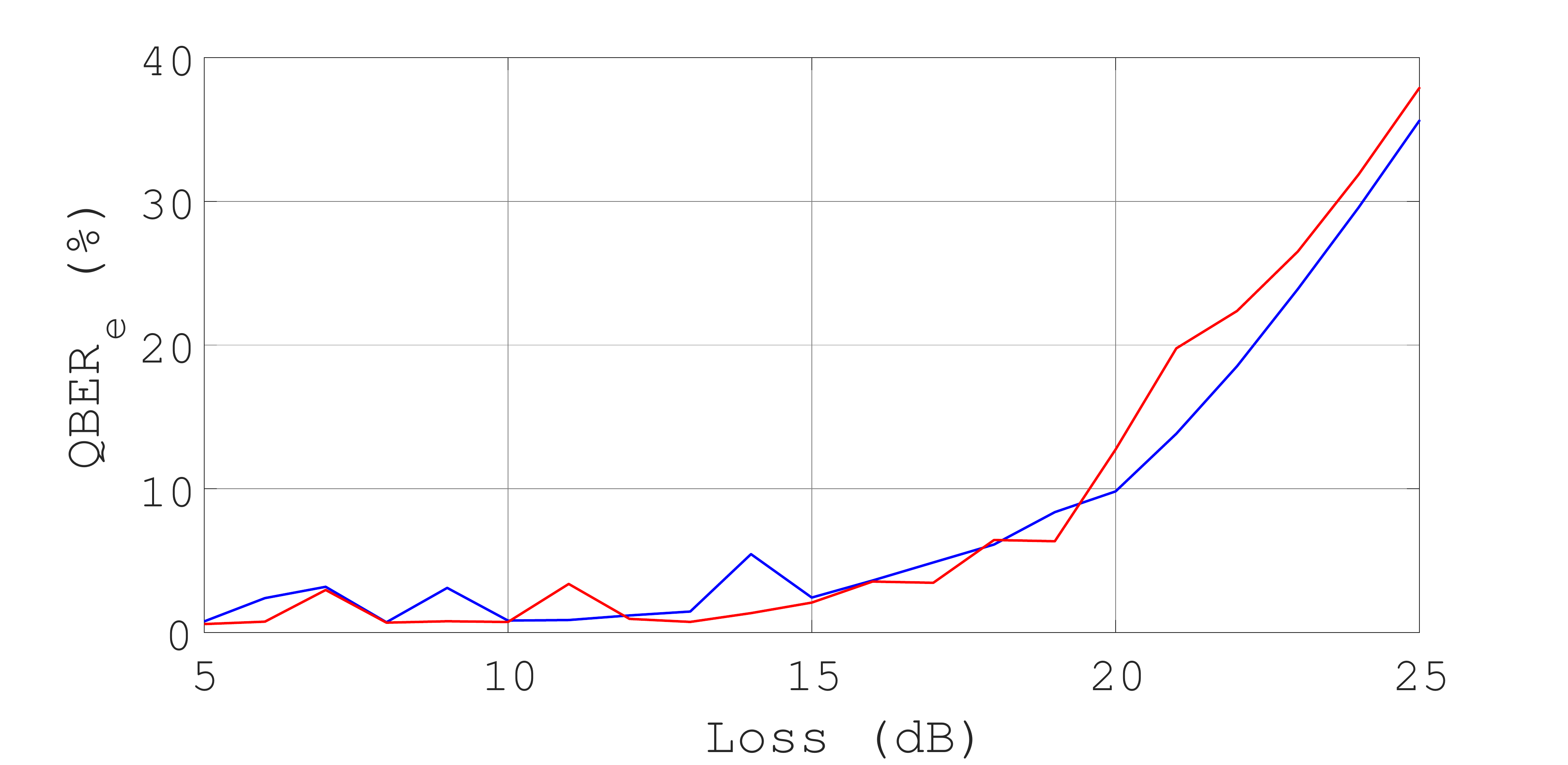}
	\caption{QBER versus line loss with Eve's improved attack. Eve can keep the error rate below 5\% for line loss upto 17 dB with the detection-scrambling-bypass strategy. The blue and red curves indicate $QBER_e$ when Bob matches $R_{ab}$ with total sifted key rate and individual channel rates respectively.} 
	\label{sec}
\end{figure}
So far, we have assumed that when Eve sends a $j$ polarized light, it is always sent towards attack angle $j$ with mean photon number $\mu_j$. In this section, we discard this assumption to generalize the attack. In particular, we assume, when Eve sends a $j$ polarized light, it can be directed towards any of the four attack angles $k \in \{ h,v,d,a\}$ with mean photon number $\mu_j^k$ and probability $f_j^k$ with $\sum_k f_j^k = 1$. Let $p_i^k(j|\theta_{B})$ be the raw click probability at Bob's detector $i$, given Eve sent a $j$-polarized light towards attack angle $k$ with mean photon number $\mu_j^k$ that has been rotated by an angle $\theta_{B}$ during scrambling. 

\begin{equation}\label{d1}
\begin{split}
p_h^k(H | 0^{\circ})  \approx c_h + 1 - exp(-\frac{\mu_H^k F \eta_k(H)}{2})
\end{split}
\end{equation}

Let $ R_e^k(j|\theta_{B})$ be the sifted key rate when Eve sends \emph{j} polarized light at \emph{k}  attack angle with Bob rotating the polarization by angle $\theta_{B}$. By deriving $R_e^k(j|\theta_{B})$ using similar analysis as \cref{eq2}-\ref{eq5} we get, 
\begin{equation}\label{eq12}
\begin{split}
R_e(H|\theta_{B}) = &[f_h^h .R_e^h(H|\theta_{B})+f_h^v.R_e^v(H|\theta_{B})+\\
&f_h^d.R_e^d(H|\theta_{B})+f_h^a.R_e^a(H|\theta_{B})]
\end{split}
\end{equation}
The above equation takes into consideration the attack angles for every polarized light sent by Eve. Thus, we now have new variables such as $P_{hv}^h(V)$ (instead of $P_{hv}(V)$) that indicates the probability -- after squashing -- that Bob selects an outcome in the $hv$ basis given the incoming light is $V$-polarized sent at $h$-attack angle. We can now plug in \cref{eq12} in \cref{equation_2} and calculate the QBER values for this detection-scrambling-bypass strategy.
In the attack model in \cite{sajeed2015a}, when Eve sent a $j$ polarized light she sent it at $j$ attack angle with mean photon number $\mu_j$ which left her with only four free parameters to minimize the error while satisfying the constraint. However, in the strategy presented in this section, when Eve decides to send a $j$ polarized light, she can send towards attack angles $k$ with probability $f_j^k$ and mean photon number $\mu_j^k$. So there are 16 different values of $\mu_j^k$ and $f_j^k$ equipping her with a total of $32$ free parameters to perform the optimization. We have solved the optimization problem for this detection-scrambling-bypass strategy with the same efficiency, Fidelity and dark count values taken from Ref.~\cite{sajeed2015a}. For matching the total rates, Eve follows the constraint $R_{ab} = R_{e}$ and for individual rates she follows $R_{ab} = R_{e}(j)$ where $j \in \{h,v,d,a \}$. With these 32 free parameters at hand the optimization program is executed and the result is shown in \cref{sec}. We see that by having more free parameters, Eve can indeed adjust their values to keep the QBER less than $5\%$ for a loss up to 17 dB.

\begin{figure}[tb]
	\centering 
a)	\includegraphics[width=0.42\columnwidth]{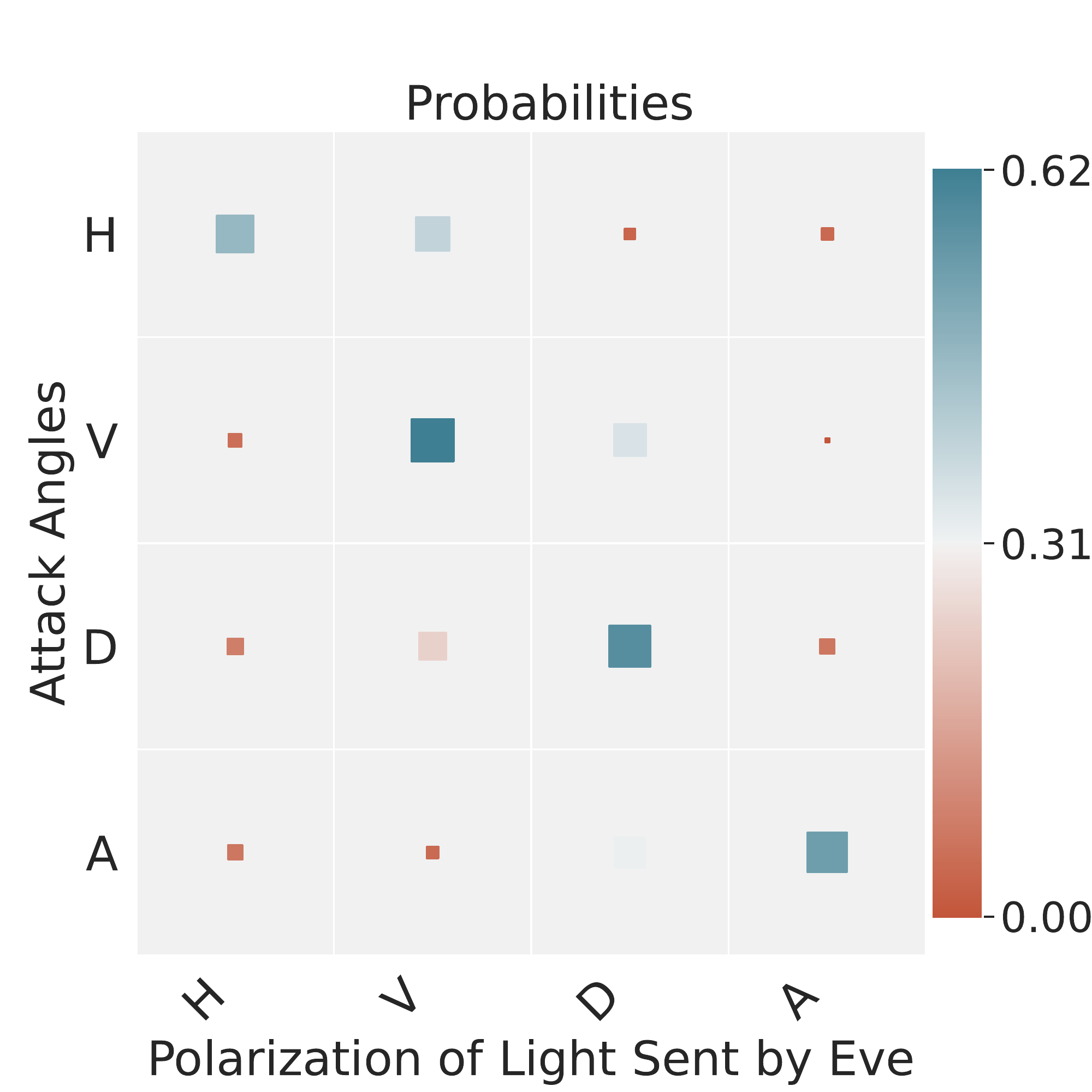} 
b)	\includegraphics[width=0.42\columnwidth]{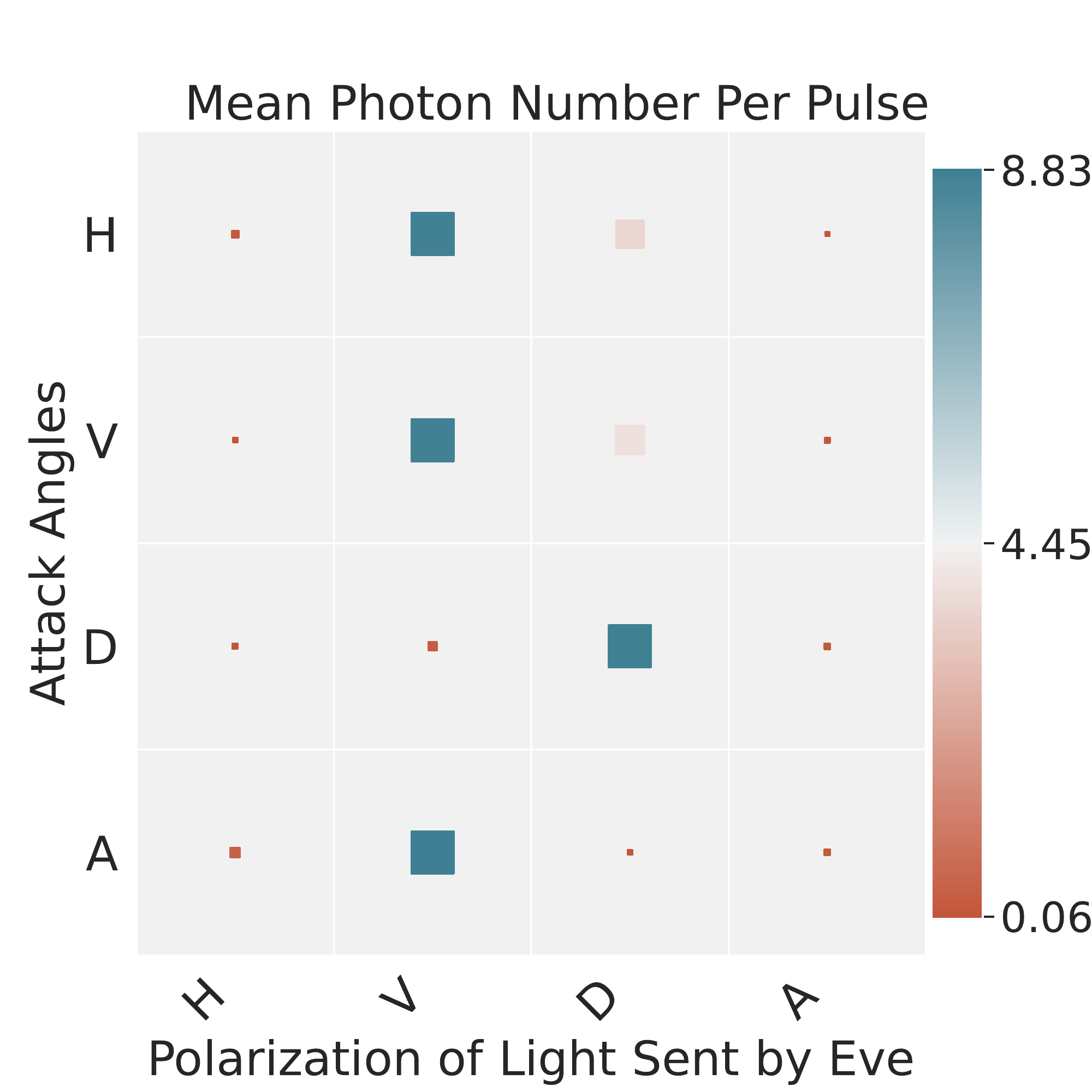}
	\caption{ a) Scatter plot of probability $f_j^k$ at channel loss of 6 dB. In the detection-scrambling-bypass strategy Eve would send a specific polarized light at all attack angles with a specific probability distribution. Each column indicates the attack angles and each row represents the polarization of light sent by Eve. We see that in most of the cases Eve sends $H$-polarized light at $h$ attack angle and so on. b) Scatter plot of mean Photon number at a channel loss of 6 dB with same column and row representation. In this case, if Eve wants to manage a successful attack, she needs to send $V$-polarized light more at $H$-attack angle than that at $V$-polarized light. Thus, depending on the window where there is total efficiency mismatch Eve needs to deploy her faked states following a specific blueprint.}
	\label{ig} 
\end{figure}

\Cref{ig}a and \cref{ig}b show the optimized probabilities $f_j^k$ and mean photon number per pulse chosen by Eve for a channel loss of 6 dB respectively. For a certain channel loss, Eve has to follow a specific blueprint to attack the system. For example, the probability plot in \cref{ig}a) shows that Eve sends \emph{V} polarized light at \emph{V} attack angle with higher probability than others. On the other hand, Eve has to send \emph{V} polarized light with higher mean photon number than other polarizations as shown in \cref{ig}b).  For different channel loss the value of the optimized free parameters will be different. Moreover, These scenarios are entirely dependent on the specific mismatch present in the system.

\section{Conclusion}
\label{conc}In this work, we have shown that randomizing the roles of the detectors cannot function as an efficient countermeasure against detector-efficiency-mismatch type attacks. Although it can prevent the original attack proposed in Ref.~\cite{sajeed2015a}, it fails to do so when a more general strategy is followed. The general strategy works even when Bob uses any non-uniform a priori scrambling probabilities. 

We note that no two practical setups will have an exact mismatch, and hence it would not be possible for Eve to acquire one prototype to learn the mismatch of the target system. However, according to Kerckhoff's principle \cite{Kerckhoffs:1883aa} quantum cryptography assumes that except for the key, Eve knows all the system's imperfections. So, to guarantee unconditional security in theory, we need to assume that Eve knows the exact details of the mismatch and Bob's scrambling countermeasure to optimize her attack. From a practical point of view, Eve can listen to Bob's classical communication channel while sending a small fraction of faked states at different spatial angles to get an estimate of the efficiency mismatch \cite{makarov2005}. Eve can pursue a similar strategy to estimate Bob's detector scrambling statistics. Thus, unless new techniques are proposed to strengthen the existing detector-scrambling countermeasure strategies, it cannot guarantee security against detector efficiency mismatch based attacks. Our result and methodology could be used to security-certify a free-space quantum communication receiver against all types of detector-efficiency-mismatch type attacks.

\bibliography{bibtex_rf}

\begin{thebibliography}{31}%
\makeatletter
\providecommand \@ifxundefined [1]{%
 \@ifx{#1\undefined}
}%
\providecommand \@ifnum [1]{%
 \ifnum #1\expandafter \@firstoftwo
 \else \expandafter \@secondoftwo
 \fi
}%
\providecommand \@ifx [1]{%
 \ifx #1\expandafter \@firstoftwo
 \else \expandafter \@secondoftwo
 \fi
}%
\providecommand \natexlab [1]{#1}%
\providecommand \enquote  [1]{``#1''}%
\providecommand \bibnamefont  [1]{#1}%
\providecommand \bibfnamefont [1]{#1}%
\providecommand \citenamefont [1]{#1}%
\providecommand \href@noop [0]{\@secondoftwo}%
\providecommand \href [0]{\begingroup \@sanitize@url \@href}%
\providecommand \@href[1]{\@@startlink{#1}\@@href}%
\providecommand \@@href[1]{\endgroup#1\@@endlink}%
\providecommand \@sanitize@url [0]{\catcode `\\12\catcode `\$12\catcode
  `\&12\catcode `\#12\catcode `\^12\catcode `\_12\catcode `\%12\relax}%
\providecommand \@@startlink[1]{}%
\providecommand \@@endlink[0]{}%
\providecommand \url  [0]{\begingroup\@sanitize@url \@url }%
\providecommand \@url [1]{\endgroup\@href {#1}{\urlprefix }}%
\providecommand \urlprefix  [0]{URL }%
\providecommand \Eprint [0]{\href }%
\providecommand \doibase [0]{http://dx.doi.org/}%
\providecommand \selectlanguage [0]{\@gobble}%
\providecommand \bibinfo  [0]{\@secondoftwo}%
\providecommand \bibfield  [0]{\@secondoftwo}%
\providecommand \translation [1]{[#1]}%
\providecommand \BibitemOpen [0]{}%
\providecommand \bibitemStop [0]{}%
\providecommand \bibitemNoStop [0]{.\EOS\space}%
\providecommand \EOS [0]{\spacefactor3000\relax}%
\providecommand \BibitemShut  [1]{\csname bibitem#1\endcsname}%
\let\auto@bib@innerbib\@empty
%</preamble>
\bibitem [{\citenamefont {Knight}(2017)}]{wknight2017}%
  \BibitemOpen
  \bibfield  {author} {\bibinfo {author} {\bibfnamefont {W.}~\bibnamefont
  {Knight}},\ }\href
  {https://www.technologyreview.com/s/609451/ibm-raises-the-bar-with-a-50-qubit-quantum-computer/?utm_campaign=Technology+Review&utm_source=facebook.com&utm_medium=social}
  {\emph {\bibinfo {title} {IBM Raises the Bar with a 50-Qubit Quantum
  Computer}}},\ \bibinfo {type} {Tech. Rep.}\ (\bibinfo  {institution} {MIT
  Technology Review},\ \bibinfo {year} {2017})\BibitemShut {NoStop}%
\bibitem [{\citenamefont {Castelvecchi}(2017)}]{dcastel}%
  \BibitemOpen
  \bibfield  {author} {\bibinfo {author} {\bibfnamefont {D.}~\bibnamefont
  {Castelvecchi}},\ }\href@noop {} {\emph {\bibinfo {title} {Quantum computers
  ready to leap out of the lab in 2017}}},\ \bibinfo {type} {Tech. Rep.}\
  (\bibinfo  {institution} {Nature News},\ \bibinfo {year} {2017})\BibitemShut
  {NoStop}%
\bibitem [{\citenamefont {Merkle}(1979)}]{stanthesis}%
  \BibitemOpen
  \bibfield  {author} {\bibinfo {author} {\bibfnamefont {R.~C.}\ \bibnamefont
  {Merkle}},\ }\emph {\bibinfo {title} {Secrecy, authentication, and public key
  systems}},\ \href@noop {} {Ph.D. thesis},\ \bibinfo  {school} {Stanford
  University} (\bibinfo {year} {1979})\BibitemShut {NoStop}%
\bibitem [{\citenamefont {Mceliece}(1978)}]{mceliece1978public}%
  \BibitemOpen
  \bibfield  {author} {\bibinfo {author} {\bibfnamefont {R.~J.}\ \bibnamefont
  {Mceliece}},\ }\href@noop {} {\bibfield  {journal} {\bibinfo  {journal}
  {Coding Thv}\ }\textbf {\bibinfo {volume} {4244}},\ \bibinfo {pages} {114}
  (\bibinfo {year} {1978})}\BibitemShut {NoStop}%
\bibitem [{\citenamefont {Hoffstein}\ \emph {et~al.}(1998)\citenamefont
  {Hoffstein}, \citenamefont {Pipher},\ and\ \citenamefont
  {Silverman}}]{hoffstein1998ntru}%
  \BibitemOpen
  \bibfield  {author} {\bibinfo {author} {\bibfnamefont {J.}~\bibnamefont
  {Hoffstein}}, \bibinfo {author} {\bibfnamefont {J.}~\bibnamefont {Pipher}}, \
  and\ \bibinfo {author} {\bibfnamefont {J.~H.}\ \bibnamefont {Silverman}},\
  }in\ \href@noop {} {\emph {\bibinfo {booktitle} {International Algorithmic
  Number Theory Symposium}}}\ (\bibinfo {organization} {Springer},\ \bibinfo
  {year} {1998})\ pp.\ \bibinfo {pages} {267--288}\BibitemShut {NoStop}%
\bibitem [{\citenamefont {Bennett}\ and\ \citenamefont
  {Brassard}(1984)}]{bennett1984}%
  \BibitemOpen
  \bibfield  {author} {\bibinfo {author} {\bibfnamefont {C.~H.}\ \bibnamefont
  {Bennett}}\ and\ \bibinfo {author} {\bibfnamefont {G.}~\bibnamefont
  {Brassard}},\ }in\ \href@noop {} {\emph {\bibinfo {booktitle} {Proceedings of
  IEEE International Conference on Computers, Systems, and Signal
  Processing}}}\ (\bibinfo  {publisher} {IEEE Press, New York},\ \bibinfo
  {address} {Bangalore, India},\ \bibinfo {year} {1984})\ pp.\ \bibinfo {pages}
  {175--179}\BibitemShut {NoStop}%
\bibitem [{\citenamefont {Gisin}\ \emph {et~al.}(2002)\citenamefont {Gisin},
  \citenamefont {Ribordy}, \citenamefont {Tittel},\ and\ \citenamefont
  {Zbinden}}]{gisin2002}%
  \BibitemOpen
  \bibfield  {author} {\bibinfo {author} {\bibfnamefont {N.}~\bibnamefont
  {Gisin}}, \bibinfo {author} {\bibfnamefont {G.}~\bibnamefont {Ribordy}},
  \bibinfo {author} {\bibfnamefont {W.}~\bibnamefont {Tittel}}, \ and\ \bibinfo
  {author} {\bibfnamefont {H.}~\bibnamefont {Zbinden}},\ }\href {\doibase
  10.1103/RevModPhys.74.145} {\bibfield  {journal} {\bibinfo  {journal} {Rev.
  Mod. Phys.}\ }\textbf {\bibinfo {volume} {74}},\ \bibinfo {pages} {145}
  (\bibinfo {year} {2002})}\BibitemShut {NoStop}%
\bibitem [{\citenamefont {Scarani}\ \emph {et~al.}(2009)\citenamefont
  {Scarani}, \citenamefont {Bechmann-Pasquinucci}, \citenamefont {Cerf},
  \citenamefont {Du\v{s}ek}, \citenamefont {L\"{u}tkenhaus},\ and\
  \citenamefont {Peev}}]{scarani2009}%
  \BibitemOpen
  \bibfield  {author} {\bibinfo {author} {\bibfnamefont {V.}~\bibnamefont
  {Scarani}}, \bibinfo {author} {\bibfnamefont {H.}~\bibnamefont
  {Bechmann-Pasquinucci}}, \bibinfo {author} {\bibfnamefont {N.~J.}\
  \bibnamefont {Cerf}}, \bibinfo {author} {\bibfnamefont {M.}~\bibnamefont
  {Du\v{s}ek}}, \bibinfo {author} {\bibfnamefont {N.}~\bibnamefont
  {L\"{u}tkenhaus}}, \ and\ \bibinfo {author} {\bibfnamefont {M.}~\bibnamefont
  {Peev}},\ }\href {\doibase 10.1103/RevModPhys.81.1301} {\bibfield  {journal}
  {\bibinfo  {journal} {Rev. Mod. Phys.}\ }\textbf {\bibinfo {volume} {81}},\
  \bibinfo {eid} {1301} (\bibinfo {year} {2009})}\BibitemShut {NoStop}%
\bibitem [{\citenamefont {Lydersen}\ \emph {et~al.}(2010)\citenamefont
  {Lydersen}, \citenamefont {Wiechers}, \citenamefont {Wittmann}, \citenamefont
  {Elser}, \citenamefont {Skaar},\ and\ \citenamefont
  {Makarov}}]{lydersen2010a}%
  \BibitemOpen
  \bibfield  {author} {\bibinfo {author} {\bibfnamefont {L.}~\bibnamefont
  {Lydersen}}, \bibinfo {author} {\bibfnamefont {C.}~\bibnamefont {Wiechers}},
  \bibinfo {author} {\bibfnamefont {C.}~\bibnamefont {Wittmann}}, \bibinfo
  {author} {\bibfnamefont {D.}~\bibnamefont {Elser}}, \bibinfo {author}
  {\bibfnamefont {J.}~\bibnamefont {Skaar}}, \ and\ \bibinfo {author}
  {\bibfnamefont {V.}~\bibnamefont {Makarov}},\ }\href {\doibase
  10.1038/nphoton.2010.214} {\bibfield  {journal} {\bibinfo  {journal} {Nat.
  Photonics}\ }\textbf {\bibinfo {volume} {4}},\ \bibinfo {pages} {686}
  (\bibinfo {year} {2010})}\BibitemShut {NoStop}%
\bibitem [{\citenamefont {Gerhardt}\ \emph {et~al.}(2011)\citenamefont
  {Gerhardt}, \citenamefont {Liu}, \citenamefont {Lamas-Linares}, \citenamefont
  {Skaar}, \citenamefont {Kurtsiefer},\ and\ \citenamefont
  {Makarov}}]{gerhardt2011}%
  \BibitemOpen
  \bibfield  {author} {\bibinfo {author} {\bibfnamefont {I.}~\bibnamefont
  {Gerhardt}}, \bibinfo {author} {\bibfnamefont {Q.}~\bibnamefont {Liu}},
  \bibinfo {author} {\bibfnamefont {A.}~\bibnamefont {Lamas-Linares}}, \bibinfo
  {author} {\bibfnamefont {J.}~\bibnamefont {Skaar}}, \bibinfo {author}
  {\bibfnamefont {C.}~\bibnamefont {Kurtsiefer}}, \ and\ \bibinfo {author}
  {\bibfnamefont {V.}~\bibnamefont {Makarov}},\ }\href {\doibase
  10.1038/ncomms1348} {\bibfield  {journal} {\bibinfo  {journal} {Nat.
  Commun.}\ }\textbf {\bibinfo {volume} {2}},\ \bibinfo {pages} {349} (\bibinfo
  {year} {2011})}\BibitemShut {NoStop}%
\bibitem [{\citenamefont {Sajeed}\ \emph
  {et~al.}(2015{\natexlab{a}})\citenamefont {Sajeed}, \citenamefont
  {Radchenko}, \citenamefont {Kaiser}, \citenamefont {Bourgoin}, \citenamefont
  {Pappa}, \citenamefont {Monat}, \citenamefont {Legr\'e},\ and\ \citenamefont
  {Makarov}}]{sajeed2015}%
  \BibitemOpen
  \bibfield  {author} {\bibinfo {author} {\bibfnamefont {S.}~\bibnamefont
  {Sajeed}}, \bibinfo {author} {\bibfnamefont {I.}~\bibnamefont {Radchenko}},
  \bibinfo {author} {\bibfnamefont {S.}~\bibnamefont {Kaiser}}, \bibinfo
  {author} {\bibfnamefont {J.-P.}\ \bibnamefont {Bourgoin}}, \bibinfo {author}
  {\bibfnamefont {A.}~\bibnamefont {Pappa}}, \bibinfo {author} {\bibfnamefont
  {L.}~\bibnamefont {Monat}}, \bibinfo {author} {\bibfnamefont
  {M.}~\bibnamefont {Legr\'e}}, \ and\ \bibinfo {author} {\bibfnamefont
  {V.}~\bibnamefont {Makarov}},\ }\href {\doibase 10.1103/PhysRevA.91.032326}
  {\bibfield  {journal} {\bibinfo  {journal} {Phys. Rev. A}\ }\textbf {\bibinfo
  {volume} {91}},\ \bibinfo {pages} {032326} (\bibinfo {year}
  {2015}{\natexlab{a}})}\BibitemShut {NoStop}%
\bibitem [{\citenamefont {Sajeed}\ \emph {et~al.}(2016)\citenamefont {Sajeed},
  \citenamefont {Huang}, \citenamefont {Sun}, \citenamefont {Xu}, \citenamefont
  {Makarov},\ and\ \citenamefont {Curty}}]{sajeed2016}%
  \BibitemOpen
  \bibfield  {author} {\bibinfo {author} {\bibfnamefont {S.}~\bibnamefont
  {Sajeed}}, \bibinfo {author} {\bibfnamefont {A.}~\bibnamefont {Huang}},
  \bibinfo {author} {\bibfnamefont {S.}~\bibnamefont {Sun}}, \bibinfo {author}
  {\bibfnamefont {F.}~\bibnamefont {Xu}}, \bibinfo {author} {\bibfnamefont
  {V.}~\bibnamefont {Makarov}}, \ and\ \bibinfo {author} {\bibfnamefont
  {M.}~\bibnamefont {Curty}},\ }\href {\doibase 10.1103/PhysRevLett.117.250505}
  {\bibfield  {journal} {\bibinfo  {journal} {Phys. Rev. Lett.}\ }\textbf
  {\bibinfo {volume} {117}},\ \bibinfo {pages} {250505} (\bibinfo {year}
  {2016})}\BibitemShut {NoStop}%
\bibitem [{\citenamefont {Jain}\ \emph {et~al.}(2014)\citenamefont {Jain},
  \citenamefont {Anisimova}, \citenamefont {Khan}, \citenamefont {Makarov},
  \citenamefont {Marquardt},\ and\ \citenamefont {Leuchs}}]{jain2014}%
  \BibitemOpen
  \bibfield  {author} {\bibinfo {author} {\bibfnamefont {N.}~\bibnamefont
  {Jain}}, \bibinfo {author} {\bibfnamefont {E.}~\bibnamefont {Anisimova}},
  \bibinfo {author} {\bibfnamefont {I.}~\bibnamefont {Khan}}, \bibinfo {author}
  {\bibfnamefont {V.}~\bibnamefont {Makarov}}, \bibinfo {author} {\bibfnamefont
  {C.}~\bibnamefont {Marquardt}}, \ and\ \bibinfo {author} {\bibfnamefont
  {G.}~\bibnamefont {Leuchs}},\ }\href {\doibase
  10.1088/1367-2630/16/12/123030} {\bibfield  {journal} {\bibinfo  {journal}
  {New J. Phys.}\ }\textbf {\bibinfo {volume} {16}},\ \bibinfo {pages} {123030}
  (\bibinfo {year} {2014})}\BibitemShut {NoStop}%
\bibitem [{\citenamefont {Sajeed}\ \emph {et~al.}(2017)\citenamefont {Sajeed},
  \citenamefont {Minshull}, \citenamefont {Jain},\ and\ \citenamefont
  {Makarov}}]{sajeed2017}%
  \BibitemOpen
  \bibfield  {author} {\bibinfo {author} {\bibfnamefont {S.}~\bibnamefont
  {Sajeed}}, \bibinfo {author} {\bibfnamefont {C.}~\bibnamefont {Minshull}},
  \bibinfo {author} {\bibfnamefont {N.}~\bibnamefont {Jain}}, \ and\ \bibinfo
  {author} {\bibfnamefont {V.}~\bibnamefont {Makarov}},\ }\href
  {https://doi.org/10.1038/s41598-017-08279-1} {\bibfield  {journal} {\bibinfo
  {journal} {Sci. Rep.}\ }\textbf {\bibinfo {volume} {7}},\ \bibinfo {pages}
  {8403} (\bibinfo {year} {2017})}\BibitemShut {NoStop}%
\bibitem [{\citenamefont {Pinheiro}\ \emph {et~al.}(2018)\citenamefont
  {Pinheiro}, \citenamefont {Chaiwongkhot}, \citenamefont {Sajeed},
  \citenamefont {Horn}, \citenamefont {Bourgoin}, \citenamefont {Jennewein},
  \citenamefont {L\"{u}tkenhaus},\ and\ \citenamefont
  {Makarov}}]{pinheiro2018}%
  \BibitemOpen
  \bibfield  {author} {\bibinfo {author} {\bibfnamefont {P.~V.~P.}\
  \bibnamefont {Pinheiro}}, \bibinfo {author} {\bibfnamefont {P.}~\bibnamefont
  {Chaiwongkhot}}, \bibinfo {author} {\bibfnamefont {S.}~\bibnamefont
  {Sajeed}}, \bibinfo {author} {\bibfnamefont {R.~T.}\ \bibnamefont {Horn}},
  \bibinfo {author} {\bibfnamefont {J.-P.}\ \bibnamefont {Bourgoin}}, \bibinfo
  {author} {\bibfnamefont {T.}~\bibnamefont {Jennewein}}, \bibinfo {author}
  {\bibfnamefont {N.}~\bibnamefont {L\"{u}tkenhaus}}, \ and\ \bibinfo {author}
  {\bibfnamefont {V.}~\bibnamefont {Makarov}},\ }\href {\doibase
  10.1364/OE.26.021020} {\bibfield  {journal} {\bibinfo  {journal} {Opt.
  Express}\ }\textbf {\bibinfo {volume} {26}},\ \bibinfo {pages} {21020}
  (\bibinfo {year} {2018})}\BibitemShut {NoStop}%
\bibitem [{\citenamefont {Makarov}\ \emph {et~al.}(2016)\citenamefont
  {Makarov}, \citenamefont {Bourgoin}, \citenamefont {Chaiwongkhot},
  \citenamefont {Gagn{\'e}}, \citenamefont {Jennewein}, \citenamefont {Kaiser},
  \citenamefont {Kashyap}, \citenamefont {Legr{\'e}}, \citenamefont
  {Minshull},\ and\ \citenamefont {Sajeed}}]{makarov2016}%
  \BibitemOpen
  \bibfield  {author} {\bibinfo {author} {\bibfnamefont {V.}~\bibnamefont
  {Makarov}}, \bibinfo {author} {\bibfnamefont {J.-P.}\ \bibnamefont
  {Bourgoin}}, \bibinfo {author} {\bibfnamefont {P.}~\bibnamefont
  {Chaiwongkhot}}, \bibinfo {author} {\bibfnamefont {M.}~\bibnamefont
  {Gagn{\'e}}}, \bibinfo {author} {\bibfnamefont {T.}~\bibnamefont
  {Jennewein}}, \bibinfo {author} {\bibfnamefont {S.}~\bibnamefont {Kaiser}},
  \bibinfo {author} {\bibfnamefont {R.}~\bibnamefont {Kashyap}}, \bibinfo
  {author} {\bibfnamefont {M.}~\bibnamefont {Legr{\'e}}}, \bibinfo {author}
  {\bibfnamefont {C.}~\bibnamefont {Minshull}}, \ and\ \bibinfo {author}
  {\bibfnamefont {S.}~\bibnamefont {Sajeed}},\ }\href {\doibase
  10.1103/PhysRevA.94.030302} {\bibfield  {journal} {\bibinfo  {journal} {Phys.
  Rev. A}\ }\textbf {\bibinfo {volume} {94}},\ \bibinfo {pages} {030302}
  (\bibinfo {year} {2016})}\BibitemShut {NoStop}%
\bibitem [{\citenamefont {Bugge}\ \emph {et~al.}(2014)\citenamefont {Bugge},
  \citenamefont {Sauge}, \citenamefont {Ghazali}, \citenamefont {Skaar},
  \citenamefont {Lydersen},\ and\ \citenamefont {Makarov}}]{bugge2014}%
  \BibitemOpen
  \bibfield  {author} {\bibinfo {author} {\bibfnamefont {A.~N.}\ \bibnamefont
  {Bugge}}, \bibinfo {author} {\bibfnamefont {S.}~\bibnamefont {Sauge}},
  \bibinfo {author} {\bibfnamefont {A.~M.~M.}\ \bibnamefont {Ghazali}},
  \bibinfo {author} {\bibfnamefont {J.}~\bibnamefont {Skaar}}, \bibinfo
  {author} {\bibfnamefont {L.}~\bibnamefont {Lydersen}}, \ and\ \bibinfo
  {author} {\bibfnamefont {V.}~\bibnamefont {Makarov}},\ }\href {\doibase
  10.1103/PhysRevLett.112.070503} {\bibfield  {journal} {\bibinfo  {journal}
  {Phys. Rev. Lett.}\ }\textbf {\bibinfo {volume} {112}},\ \bibinfo {pages}
  {070503} (\bibinfo {year} {2014})}\BibitemShut {NoStop}%
\bibitem [{\citenamefont {Bennett}\ \emph {et~al.}(1992)\citenamefont
  {Bennett}, \citenamefont {Bessette}, \citenamefont {Salvail}, \citenamefont
  {Brassard},\ and\ \citenamefont {Smolin}}]{bennett1992b}%
  \BibitemOpen
  \bibfield  {author} {\bibinfo {author} {\bibfnamefont {C.~H.}\ \bibnamefont
  {Bennett}}, \bibinfo {author} {\bibfnamefont {F.}~\bibnamefont {Bessette}},
  \bibinfo {author} {\bibfnamefont {L.}~\bibnamefont {Salvail}}, \bibinfo
  {author} {\bibfnamefont {G.}~\bibnamefont {Brassard}}, \ and\ \bibinfo
  {author} {\bibfnamefont {J.}~\bibnamefont {Smolin}},\ }\href {\doibase
  10.1007/bf00191318} {\bibfield  {journal} {\bibinfo  {journal} {J.
  Cryptology}\ }\textbf {\bibinfo {volume} {5}},\ \bibinfo {pages} {3}
  (\bibinfo {year} {1992})}\BibitemShut {NoStop}%
\bibitem [{\citenamefont {Xu}\ \emph {et~al.}(2015)\citenamefont {Xu},
  \citenamefont {Wei}, \citenamefont {Sajeed}, \citenamefont {Kaiser},
  \citenamefont {Sun}, \citenamefont {Tang}, \citenamefont {Qian},
  \citenamefont {Makarov},\ and\ \citenamefont {Lo}}]{xu2015a}%
  \BibitemOpen
  \bibfield  {author} {\bibinfo {author} {\bibfnamefont {F.}~\bibnamefont
  {Xu}}, \bibinfo {author} {\bibfnamefont {K.}~\bibnamefont {Wei}}, \bibinfo
  {author} {\bibfnamefont {S.}~\bibnamefont {Sajeed}}, \bibinfo {author}
  {\bibfnamefont {S.}~\bibnamefont {Kaiser}}, \bibinfo {author} {\bibfnamefont
  {S.}~\bibnamefont {Sun}}, \bibinfo {author} {\bibfnamefont {Z.}~\bibnamefont
  {Tang}}, \bibinfo {author} {\bibfnamefont {L.}~\bibnamefont {Qian}}, \bibinfo
  {author} {\bibfnamefont {V.}~\bibnamefont {Makarov}}, \ and\ \bibinfo
  {author} {\bibfnamefont {H.-K.}\ \bibnamefont {Lo}},\ }\href {\doibase
  10.1103/PhysRevA.92.032305} {\bibfield  {journal} {\bibinfo  {journal} {Phys.
  Rev. A}\ }\textbf {\bibinfo {volume} {92}},\ \bibinfo {pages} {032305}
  (\bibinfo {year} {2015})}\BibitemShut {NoStop}%
\bibitem [{\citenamefont {Makarov}\ \emph {et~al.}(2006)\citenamefont
  {Makarov}, \citenamefont {Anisimov},\ and\ \citenamefont
  {Skaar}}]{makarov2006}%
  \BibitemOpen
  \bibfield  {author} {\bibinfo {author} {\bibfnamefont {V.}~\bibnamefont
  {Makarov}}, \bibinfo {author} {\bibfnamefont {A.}~\bibnamefont {Anisimov}}, \
  and\ \bibinfo {author} {\bibfnamefont {J.}~\bibnamefont {Skaar}},\ }\href
  {\doibase 10.1103/PhysRevA.74.022313} {\bibfield  {journal} {\bibinfo
  {journal} {Phys. Rev. A}\ }\textbf {\bibinfo {volume} {74}},\ \bibinfo
  {pages} {022313} (\bibinfo {year} {2006})},\ \bibinfo {note} {erratum ibid.
  \textbf{78}, 019905 (2008)}\BibitemShut {NoStop}%
\bibitem [{\citenamefont {Sajeed}\ \emph
  {et~al.}(2015{\natexlab{b}})\citenamefont {Sajeed}, \citenamefont
  {Chaiwongkhot}, \citenamefont {Bourgoin}, \citenamefont {Jennewein},
  \citenamefont {L{\"u}tkenhaus},\ and\ \citenamefont {Makarov}}]{sajeed2015a}%
  \BibitemOpen
  \bibfield  {author} {\bibinfo {author} {\bibfnamefont {S.}~\bibnamefont
  {Sajeed}}, \bibinfo {author} {\bibfnamefont {P.}~\bibnamefont
  {Chaiwongkhot}}, \bibinfo {author} {\bibfnamefont {J.-P.}\ \bibnamefont
  {Bourgoin}}, \bibinfo {author} {\bibfnamefont {T.}~\bibnamefont {Jennewein}},
  \bibinfo {author} {\bibfnamefont {N.}~\bibnamefont {L{\"u}tkenhaus}}, \ and\
  \bibinfo {author} {\bibfnamefont {V.}~\bibnamefont {Makarov}},\ }\href
  {\doibase 10.1103/PhysRevA.91.062301} {\bibfield  {journal} {\bibinfo
  {journal} {Phys. Rev. A}\ }\textbf {\bibinfo {volume} {91}},\ \bibinfo
  {pages} {062301} (\bibinfo {year} {2015}{\natexlab{b}})}\BibitemShut
  {NoStop}%
\bibitem [{\citenamefont {Rau}\ \emph {et~al.}(2015)\citenamefont {Rau},
  \citenamefont {Vogl}, \citenamefont {Corrielli}, \citenamefont {Vest},
  \citenamefont {Fuchs}, \citenamefont {Nauerth},\ and\ \citenamefont
  {Weinfurter}}]{rau2015}%
  \BibitemOpen
  \bibfield  {author} {\bibinfo {author} {\bibfnamefont {M.}~\bibnamefont
  {Rau}}, \bibinfo {author} {\bibfnamefont {T.}~\bibnamefont {Vogl}}, \bibinfo
  {author} {\bibfnamefont {G.}~\bibnamefont {Corrielli}}, \bibinfo {author}
  {\bibfnamefont {G.}~\bibnamefont {Vest}}, \bibinfo {author} {\bibfnamefont
  {L.}~\bibnamefont {Fuchs}}, \bibinfo {author} {\bibfnamefont
  {S.}~\bibnamefont {Nauerth}}, \ and\ \bibinfo {author} {\bibfnamefont
  {H.}~\bibnamefont {Weinfurter}},\ }\href {\doibase
  10.1109/JSTQE.2014.2372008} {\bibfield  {journal} {\bibinfo  {journal} {IEEE
  J. Quantum. Electron.}\ }\textbf {\bibinfo {volume} {21}},\ \bibinfo {pages}
  {6600905} (\bibinfo {year} {2015})}\BibitemShut {NoStop}%
\bibitem [{\citenamefont {Chaiwongkhot}\ \emph {et~al.}(2019)\citenamefont
  {Chaiwongkhot}, \citenamefont {Kuntz}, \citenamefont {Zhang}, \citenamefont
  {Huang}, \citenamefont {Bourgoin}, \citenamefont {Sajeed}, \citenamefont
  {L\"utkenhaus}, \citenamefont {Jennewein},\ and\ \citenamefont
  {Makarov}}]{chaiwongkhot2019}%
  \BibitemOpen
  \bibfield  {author} {\bibinfo {author} {\bibfnamefont {P.}~\bibnamefont
  {Chaiwongkhot}}, \bibinfo {author} {\bibfnamefont {K.~B.}\ \bibnamefont
  {Kuntz}}, \bibinfo {author} {\bibfnamefont {Y.}~\bibnamefont {Zhang}},
  \bibinfo {author} {\bibfnamefont {A.}~\bibnamefont {Huang}}, \bibinfo
  {author} {\bibfnamefont {J.-P.}\ \bibnamefont {Bourgoin}}, \bibinfo {author}
  {\bibfnamefont {S.}~\bibnamefont {Sajeed}}, \bibinfo {author} {\bibfnamefont
  {N.}~\bibnamefont {L\"utkenhaus}}, \bibinfo {author} {\bibfnamefont
  {T.}~\bibnamefont {Jennewein}}, \ and\ \bibinfo {author} {\bibfnamefont
  {V.}~\bibnamefont {Makarov}},\ }\href {\doibase 10.1103/PhysRevA.99.062315}
  {\bibfield  {journal} {\bibinfo  {journal} {Phys. Rev. A}\ }\textbf {\bibinfo
  {volume} {99}},\ \bibinfo {pages} {062315} (\bibinfo {year}
  {2019})}\BibitemShut {NoStop}%
\bibitem [{\citenamefont {Qi}\ \emph {et~al.}(2007)\citenamefont {Qi},
  \citenamefont {Fung}, \citenamefont {Lo},\ and\ \citenamefont {Ma}}]{qi2007}%
  \BibitemOpen
  \bibfield  {author} {\bibinfo {author} {\bibfnamefont {B.}~\bibnamefont
  {Qi}}, \bibinfo {author} {\bibfnamefont {C.-H.~F.}\ \bibnamefont {Fung}},
  \bibinfo {author} {\bibfnamefont {H.-K.}\ \bibnamefont {Lo}}, \ and\ \bibinfo
  {author} {\bibfnamefont {X.}~\bibnamefont {Ma}},\ }\href@noop {} {\bibfield
  {journal} {\bibinfo  {journal} {Quant. Inf. Comp.}\ }\textbf {\bibinfo
  {volume} {7}},\ \bibinfo {pages} {73} (\bibinfo {year} {2007})}\BibitemShut
  {NoStop}%
\bibitem [{\citenamefont {Makarov}\ and\ \citenamefont
  {Skaar}(2008)}]{makarov2008}%
  \BibitemOpen
  \bibfield  {author} {\bibinfo {author} {\bibfnamefont {V.}~\bibnamefont
  {Makarov}}\ and\ \bibinfo {author} {\bibfnamefont {J.}~\bibnamefont
  {Skaar}},\ }\href@noop {} {\bibfield  {journal} {\bibinfo  {journal} {Quant.
  Inf. Comp.}\ }\textbf {\bibinfo {volume} {8}},\ \bibinfo {pages} {622}
  (\bibinfo {year} {2008})}\BibitemShut {NoStop}%
\bibitem [{\citenamefont {da~Silva}\ \emph {et~al.}(2014)\citenamefont
  {da~Silva}, \citenamefont {do~Amaral}, \citenamefont {Xavier}, \citenamefont
  {Tempor{\~a}o},\ and\ \citenamefont {von~der Weid}}]{da2014safeguarding}%
  \BibitemOpen
  \bibfield  {author} {\bibinfo {author} {\bibfnamefont {T.~F.}\ \bibnamefont
  {da~Silva}}, \bibinfo {author} {\bibfnamefont {G.~C.}\ \bibnamefont
  {do~Amaral}}, \bibinfo {author} {\bibfnamefont {G.~B.}\ \bibnamefont
  {Xavier}}, \bibinfo {author} {\bibfnamefont {G.~P.}\ \bibnamefont
  {Tempor{\~a}o}}, \ and\ \bibinfo {author} {\bibfnamefont {J.~P.}\
  \bibnamefont {von~der Weid}},\ }\href@noop {} {\bibfield  {journal} {\bibinfo
   {journal} {IEEE Journal of Selected Topics in Quantum Electronics}\ }\textbf
  {\bibinfo {volume} {21}},\ \bibinfo {pages} {159} (\bibinfo {year}
  {2014})}\BibitemShut {NoStop}%
\bibitem [{\citenamefont {Kerckhoffs}(1883)}]{Kerckhoffs:1883aa}%
  \BibitemOpen
  \bibfield  {author} {\bibinfo {author} {\bibfnamefont {A.}~\bibnamefont
  {Kerckhoffs}},\ }\href
  {https://www.petitcolas.net/kerckhoffs/crypto_militaire_1_b.pdf} {\bibfield
  {journal} {\bibinfo  {journal} {Journal des sciences militaires}\ }\textbf
  {\bibinfo {volume} {IX}},\ \bibinfo {pages} {5} (\bibinfo {year}
  {1883})}\BibitemShut {NoStop}%
\bibitem [{\citenamefont {Makarov}\ and\ \citenamefont
  {Hjelme}(2005)}]{makarov2005}%
  \BibitemOpen
  \bibfield  {author} {\bibinfo {author} {\bibfnamefont {V.}~\bibnamefont
  {Makarov}}\ and\ \bibinfo {author} {\bibfnamefont {D.~R.}\ \bibnamefont
  {Hjelme}},\ }\href {\doibase 10.1080/09500340410001730986} {\bibfield
  {journal} {\bibinfo  {journal} {J. Mod. Opt.}\ }\textbf {\bibinfo {volume}
  {52}},\ \bibinfo {pages} {691} (\bibinfo {year} {2005})}\BibitemShut
  {NoStop}%
\bibitem [{\citenamefont {Beaudry}\ \emph {et~al.}(2008)\citenamefont
  {Beaudry}, \citenamefont {Moroder},\ and\ \citenamefont
  {L\"{u}tkenhaus}}]{beaudry2008}%
  \BibitemOpen
  \bibfield  {author} {\bibinfo {author} {\bibfnamefont {N.~J.}\ \bibnamefont
  {Beaudry}}, \bibinfo {author} {\bibfnamefont {T.}~\bibnamefont {Moroder}}, \
  and\ \bibinfo {author} {\bibfnamefont {N.}~\bibnamefont {L\"{u}tkenhaus}},\
  }\href {\doibase 10.1103/PhysRevLett.101.093601} {\bibfield  {journal}
  {\bibinfo  {journal} {Phys. Rev. Lett.}\ }\textbf {\bibinfo {volume} {101}},\
  \bibinfo {eid} {093601} (\bibinfo {year} {2008})}\BibitemShut {NoStop}%
\bibitem [{\citenamefont {Tsurumaru}\ and\ \citenamefont
  {Tamaki}(2008)}]{tsurumaru2008}%
  \BibitemOpen
  \bibfield  {author} {\bibinfo {author} {\bibfnamefont {T.}~\bibnamefont
  {Tsurumaru}}\ and\ \bibinfo {author} {\bibfnamefont {K.}~\bibnamefont
  {Tamaki}},\ }\href {\doibase 10.1103/PhysRevA.78.032302} {\bibfield
  {journal} {\bibinfo  {journal} {Phys. Rev. A}\ }\textbf {\bibinfo {volume}
  {78}},\ \bibinfo {eid} {032302} (\bibinfo {year} {2008})}\BibitemShut
  {NoStop}%
\bibitem [{\citenamefont {Gittsovich}\ \emph {et~al.}(2014)\citenamefont
  {Gittsovich}, \citenamefont {Beaudry}, \citenamefont {Narasimhachar},
  \citenamefont {Alvarez}, \citenamefont {Moroder},\ and\ \citenamefont
  {L{\"u}tkenhaus}}]{gittsovich2014}%
  \BibitemOpen
  \bibfield  {author} {\bibinfo {author} {\bibfnamefont {O.}~\bibnamefont
  {Gittsovich}}, \bibinfo {author} {\bibfnamefont {N.~J.}\ \bibnamefont
  {Beaudry}}, \bibinfo {author} {\bibfnamefont {V.}~\bibnamefont
  {Narasimhachar}}, \bibinfo {author} {\bibfnamefont {R.~R.}\ \bibnamefont
  {Alvarez}}, \bibinfo {author} {\bibfnamefont {T.}~\bibnamefont {Moroder}}, \
  and\ \bibinfo {author} {\bibfnamefont {N.}~\bibnamefont {L{\"u}tkenhaus}},\
  }\href {\doibase 10.1103/PhysRevA.89.012325} {\bibfield  {journal} {\bibinfo
  {journal} {Phys. Rev. A}\ }\textbf {\bibinfo {volume} {89}},\ \bibinfo
  {pages} {012325} (\bibinfo {year} {2014})}\BibitemShut {NoStop}%
\end{thebibliography}%

\appendix
\section{Sifted key rate and QBER during attack}
\label{squashing}

To derive the key rate and QBER formula in Eve's presence, Ref~\cite{sajeed2015a} started with a system with only Eve and Bob. Let us consider Eve is sending a $j$-polarized pulse to Bob with mean photon number $\mu_j$ towards the attack angles $j$. Let $p_i(j)$ be the raw click probability at detector $i$ while incoming light is $j$ polarized. For Eve sending $H$ polarized light, these probabilities are:
\begin{equation}
\label{rcb}
\begin{aligned}
& p_h(H)  \approx  ~c_h + 1 - \exp(-\frac{\mu_H F \eta_h(H)}{2})\\
& p_v(H)  \approx  ~c_v + 1 - \exp(-\frac{\mu_H (1-F) \eta_v(H)}{2})\\
& p_{d(a)}(H)  \approx  ~c_{d(a)} + 1 - \exp(-\frac{\mu_H \eta_{d(a)}(H)}{4})
\end{aligned}
\end{equation}

Here, $c_i$ is the dark count probability per bit slot at the $i$-th detector, F is the fidelity and $\eta_i(j)$ is the probability of detection at Bob's $i$-th detector given Eve sent $j$-polarized light. We assume that when Bob registers a multiple click, he performs a squashing operation\cite{beaudry2008,tsurumaru2008,gittsovich2014}. \cref{tab:table1} shows the cases where Bob makes his decisions based on the clicks on his detectors. All other cases are discarded in the squashing model. Let $\mathrm{P_{k}(l)}$ (where $k \in \{hv,da\}$, $l \in \{H,V,D,A\}$) be the probability that Bob measures in the $k$ basis given the incoming light is $l$ polarized. Then $P_{hv}(H)$ can be computed from cases $1$, $2$ and $5$ in \cref{tab:table1} as: 
\begin{equation} \label{phvi}
\begin{split}
P_{hv}(H) &= p_h(H)[1 - p_d(H)][1 - p_a(H)][1 - p_v(H)]\\
& \quad + p_v(H)[1 - p_d(H)][1 - p_a(H)][1 - p_h(H)]\\
& \quad + p_v(H)p_h(H)[1 - p_d(H)][1 - p_a(H)]\\
& = [1 - p_d(H)][1 - p_a(H)]\\
& \quad \times [p_h(H) + p_v(H) - p_h(H)p_v(H)].
\end{split}
\end{equation}

\begin{table}[tb]
	\begin{center}
		\caption{Possible outcome of events after squashing.}
		\label{tab:table1}
		\begin{tabular}{c|c|c|c|c|c} % <-- Alignments: 1st column left, 2nd middle and 3rd right, with vertical lines in between
			\text{Case \#}& \textbf{$h$} & \textbf{$v$} & \textbf{$d$} & \textbf{$a$} & \text{Decision after squashing} \\
			%$\alpha$ & $\beta$ & $\gamma$ \\
			\hline 1 & $\checkmark$ & $\times$ & $\times$ & $\times$ & click on $h$\\
			\hline  2 & $\times$ & $\checkmark$ & $\times$ & $\times$ & click on $v$\\
			\hline 3 & $\times$ & $\times$ & $\checkmark$ & $\times$ & click on $d$\\
			\hline 4 & $\times$ & $\times$ & $\times$ & $\checkmark$ & click on $a$\\
			\hline 5 & $\checkmark$ & $\checkmark$ & $\times$ & $\times$ & decision random $h$ or $v$\\
			\hline  6 & $\times$ & $\times$ & $\checkmark$ & $\checkmark$ & decision random $d$ or $a$
		\end{tabular}
	\end{center}
\end{table}

\begin{figure}[htb]
	\centering
	\includegraphics[width=80mm]{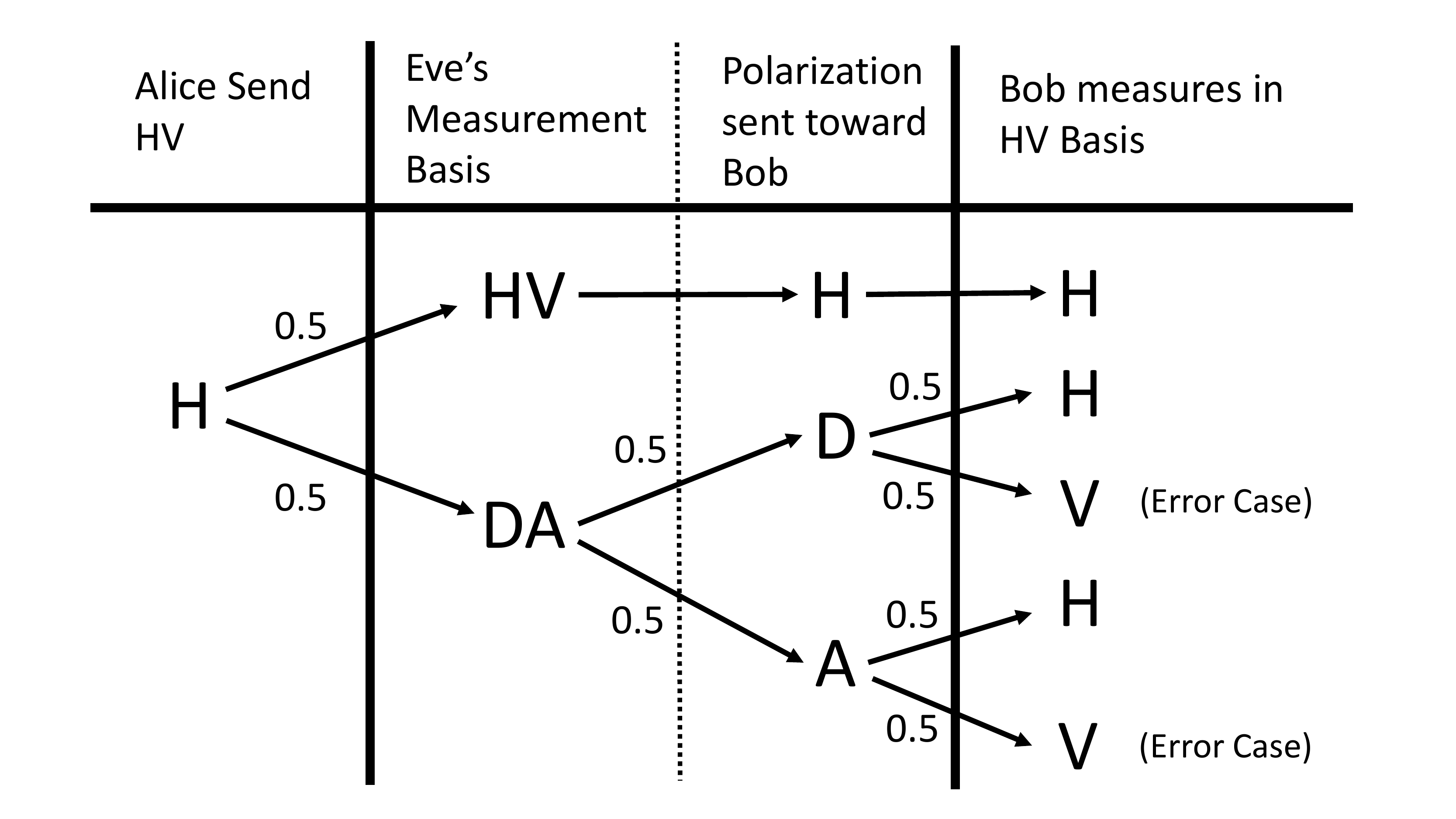}
	\caption{A Bayesian network showing the possible scenarios if Alice sends $H$-polarized light and Bob measure in $HV$-basis. This only portrays a scenario where fidelity is 1.0 with no dark counts.}
	\label{net}
\end{figure}

\noindent The probabilities $P_{hv}(V),P_{da}(D),P_{da}(A)$ can be calculated similarly. Now we include Alice into the picture. We first assume the case where Alice sends a $H$-polarized light.  The possible scenarios are shown in \cref{net}. It is sufficient to consider only the cases when Bob measures in same basis as Alice (HV in this case) as the other cases will be discarded during sifting. Here we assume, Eve measures Alice's outgoing signal in $HV$ or $DA$ basis with equal a-priory probability using a measurement setup having perfect detection efficiency and no dark count. Thus, with $50\%$ probability she measures in the correct (incorrect) basis and sends the correct (incorrect) state to Bob.   Let $R_e(j)$ be the sifted key rate with Eve's presence given Alice sent a $j$ polarized light. Following \cref{net}, $R_e(j)$ can be given by, 
\begin{equation} \label{reh}
R_e(H) \approx \frac{1}{2} P_{hv}(H) + \frac{1}{4} P_{hv}(D) + \frac{1}{4} P_{hv}(A)
\end{equation}
The error rate with Eve given Alice sends a $H$ polarized light can also be calculated with the help of \cref{net}. When Eve measures in the same basis as Alice, she introduces no error (assuming perfect fidelity at Bob). However, when she measures in the wrong basis (in this case, $DA$) there is some probability of error. Let $\mathrm{P_i(j)}$ be the probability that, after squashing, Bob decides on outcome $i$  given incoming light was $j$-polarized light. Thus, $\mathrm{P_v(H)}$ would be, 
\begin{equation} \label{pvhp}
\mathrm{P_v(H)} = [p_v(H) - \frac{p_h(H) p_v(H)}{2}][1 - p_d(H)][1-p_a(H)]
\end{equation}

Hence, the error rate during attack given Alice sends a $H$-polarized light is,
\begin{equation} 
\label{serror}
E_H = \frac{1}{8} \mathrm{P_v(D)} + \frac{1}{8} \mathrm{P_v(A)}
\end{equation}

In deriving  \cref{reh,pvhp,serror}, we have assumed simplified cases. In a more general scenario, we also need to consider $P_{hv}(V)$ since the setup may have imperfect fidelity and dark counts in the photodetectors. Let $P_c^e$ and $P_w^e$ be the probability that Eve measures Alice's signal in the correct basis and gets a click in the correct and wrong photodetector respectively. Let, $P_{nc}^e$ be the probability that Eve measures in the non-compatible or wrong basis. We can then modify equation \ref{reh} for the case of sifted key rate when there is incoming $H$-polarized light. Thus, the sifted key rate can be written from \cite{sajeed2015a} in the following form

\begin{equation} \label{rate}
\begin{split}
R_e(H) \approx &P_c^e P_{hv}(H) + P_w^e P_{hv}(V) + P_{nc}^e [ P_{hv}(D) + P_{hv}(A)]\\
& + (1 - P_c^e - P_w^e - 2P_{nc}^e) (c_h + c_v - c_h c_v)
\end{split}
\end{equation}
Similarly, we can modify equation \ref{serror} to calculate the error rate with Eve in between, when Alice sends $H$-polarized light. If Bob has a click in the $v$ photodetector with incoming $H$-polarized light then that would be an error case with Eve measuring Alice's signal in the correct basis. Similarly, for other cases where Eve measures in the correct basis but gets a click in the wrong photodetector and Eve measuring in the wrong basis, if Bob gets a click in the $v$ photodetector that would count as an error and can be expressed in the following form:
\begin{equation} \label{error}
\begin{split}
E_H \approx &P_c^e \mathrm{P_v(H)} + P_w^e \mathrm{P_v(V)} + P_{nc}^e [\mathrm{P_v(D)} + \mathrm{P_v(A)}]\\
& + (1 - P_c^e - P_w^e - 2 P_{nc}^e) (c_v - \frac{c_v c_h}{2})
\end{split}
\end{equation}

Sifted key rates and QBERs during attack given Alice sends $V$, $D$ and $A$ polarized light can be calculated similarly. The total sifted key rate and QBER in Eve's presence become
\begin{equation}
\begin{aligned}
&R_e = \frac{1}{4} \sum_{j = H,V,D,A} R_e(j),\\
&\text{QBER}_e =\frac{1}{4 R_e} \sum_{j = H,V,D,A} E_j.
\end{aligned}
\end{equation}

\section{Sifted key rate and QBER with scrambling countermeasure}
\label{scrambling}

\begin{figure}[tb]
	\centering
	\includegraphics[width=0.7\columnwidth]{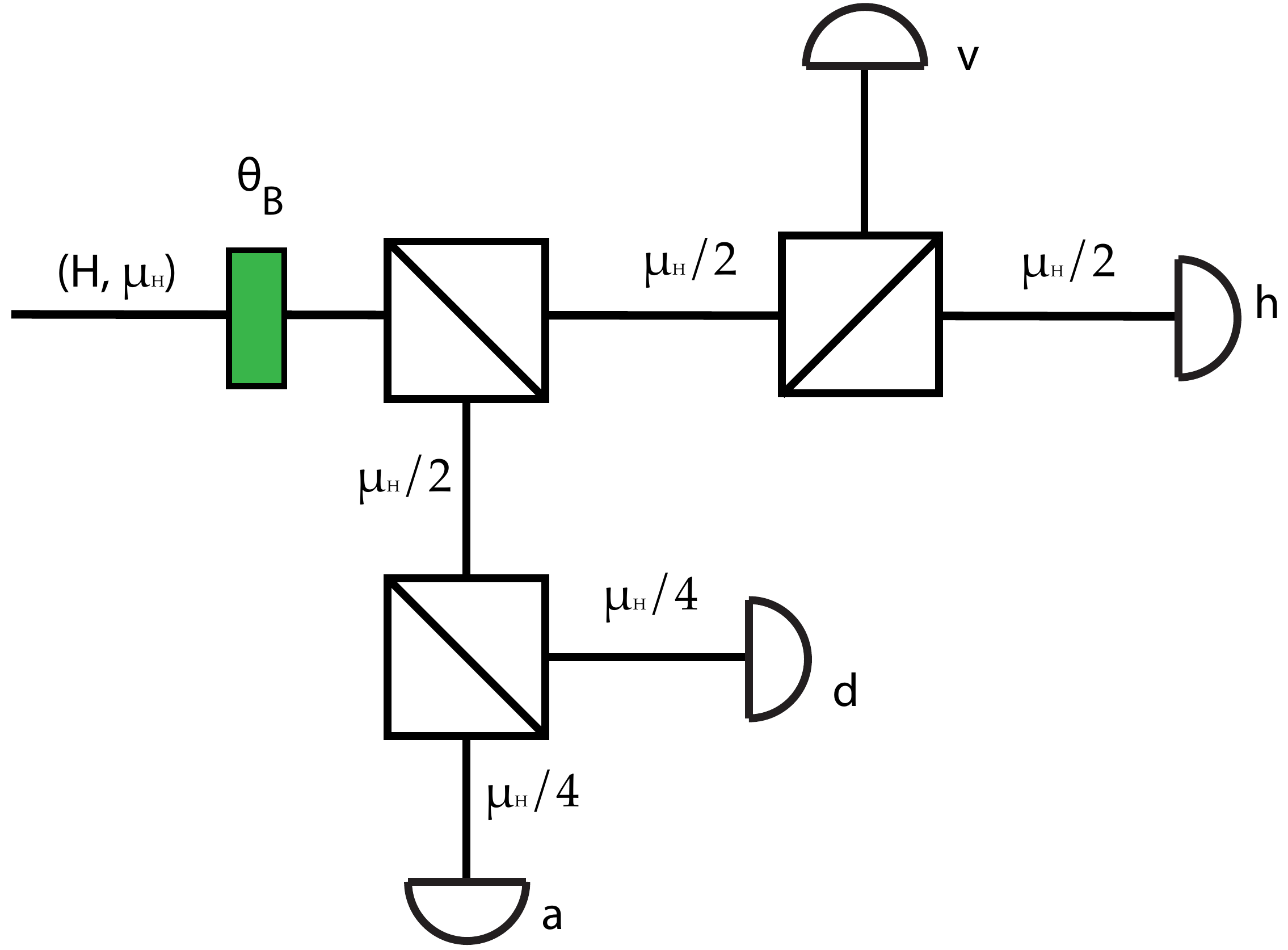}
	\caption{Bob's measurement setup during scrambling}
\end{figure}

Let $p_i(j | \theta_B)$ be the raw click probability  at Bob's  $i$-th detector given Eve sends $j$ polarized light with mean photon number $\mu_j$ directed towards attack angle $j$ which is rotated by Bob by an angle $\theta_{B}$. The probabilities for $\theta_B = 0\degree, 45~\degree, 90~\degree$ and $135~\degree$ can be derived similar to \cref{rcb}. When $\theta_B = 0~\degree$:

\begin{equation}
\begin{aligned}
& p_h(H | 0^{\circ})  \approx  ~c_h + 1 - \exp(-\frac{\mu_H F \eta_h(H)}{2})\\
& p_v(H | 0^{\circ})  \approx  ~c_v + 1 - \exp(-\frac{\mu_H (1-F) \eta_v(H)}{2})\\
&p_{d(a)}(H | 0^{\circ})  \approx  ~c_{d(a)} + 1 - \exp(-\frac{\mu_H \eta_{d(a)}(H)}{4})
\end{aligned}
\end{equation}

When $\theta_B = 45~\degree $ the $H$-polarized light is rotated to a $D$-polarized light and corresponding raw click probabilities become:

\begin{equation}
\begin{aligned}
&p_d(H | 45^{\circ})  \approx c_d + 1 - exp(-\frac{\mu_H F \eta_d(H)}{2}) \\
&p_a(H | 45^{\circ})  \approx c_a + 1 - exp(-\frac{\mu_H (1-F) \eta_a(H)}{2}) \\
&p_{h(v)}(H | 45^{\circ})  \approx c_{h(v)} + 1 - exp(-\frac{\mu_H \eta_{h(v)}(H)}{4})
\end{aligned}
\end{equation}
\noindent For  $\theta_{B} = 90~\degree$:
\begin{equation}
\begin{aligned}
&p_v(H | 90^{\circ})  \approx c_v + 1 - exp(-\frac{\mu_H F \eta_v(H)}{2})\\
&p_h(H | 90^{\circ})  \approx c_h + 1 - exp(-\frac{\mu_H (1-F) \eta_h(H)}{2})\\
&p_{d(a)}(H | 90^{\circ})  \approx c_{d(a)} + 1 - exp(-\frac{\mu_H \eta_{d(a)}(H)}{4})
\end{aligned}
\end{equation}
\noindent and finally for $\theta_B = 135~\degree$:
\begin{equation}
\begin{aligned}
&p_a(H | 135^{\circ})  \approx c_a + 1 - exp(-\frac{\mu_H F \eta_a(H)}{2}) \\
&p_d(H | 135^{\circ})  \approx c_d + 1 - exp(-\frac{\mu_H (1-F) \eta_d(H)}{2})\\
&p_{h(v)}(H | 135^{\circ})  \approx c_{h(v)} + 1 - exp(-\frac{\mu_H \eta_{h(v)}(H)}{4})
\end{aligned}
\end{equation}

Let $R_e(j|\theta_B)$ be the sifted key rate in the presence of Eve given she sends $j$ polarized light -- towards attack angle $j$ with mean photon number $\mu_j$ --  and Bob applies $\theta_B$ rotation on it.  Using similar analysis used for deriving \cref{rate}, we can find the rates for different $\theta_B$. For example, 
\begin{equation} \label{eq2}
\begin{split}
R_e(H|&0^{\circ}) \approx P_{c}^{e}P_{hv}(H|0^{\circ}) + P_{w}^{e}P_{hv}(V |0^{\circ})\\
& + P_{nc}^{e}[P_{hv}(D|0^{\circ}) + P_{hv}(A | 0^{\circ})]\\
&  + (1 - P_{c}^{e} - P_{w}^{e} - 2P_{nc}^{e}) (c_{h} + c_{v} - c_{h} c_{v}) 
\end{split}
\end{equation}

\begin{equation} \label{eq3}
\begin{split}
R_e(H|&90^{\circ}) \approx P_{c}^{e}P_{hv}(H|90^{\circ}) + P_{w}^{e}P_{hv}(V |90^{\circ})\\
& + P_{nc}^{e}[P_{hv}(D|90^{\circ}) + P_{hv}(A | 90^{\circ})]\\
&  + (1 - P_{c}^{e} - P_{w}^{e} - 2P_{nc}^{e}) (c_{h} + c_{v} - c_{h} c_{v}) 
\end{split}
\end{equation}

\begin{equation} \label{eq4}
\begin{split}
R_e(H|&45^{\circ}) \approx P_{c}^{e}P_{da}(H|45^{\circ}) + P_{w}^{e}P_{da}(V |45^{\circ})\\
& + P_{nc}^{e}[P_{da}(D|45^{\circ}) + P_{da}(A | 45^{\circ})]\\
&  + (1 - P_{c}^{e} - P_{w}^{e} - 2P_{nc}^{e}) (c_{d} + c_{a} - c_{d} c_{a}) 
\end{split}
\end{equation}

\begin{equation} \label{eq5}
\begin{split}
R_e(H|&135^{\circ}) \approx P_{c}^{e}P_{da}(H |135^{\circ}) + P_{w}^{e}P_{da}(V |135^{\circ})\\
& + P_{nc}^{e}[P_{da}(D|135^{\circ}) + P_{da}(A | 135^{\circ})]\\
&  + (1 - P_{c}^{e} - P_{w}^{e} - 2P_{nc}^{e}) (c_{d} + c_{a} - c_{d} c_{a}) 
\end{split}
\end{equation}

We assume in our model that Bob is scrambling the role of the photodetectors with equal a-priori probabilities. Thus, modifying equation \ref{r_qber} and averaging Bob's rate for each $\theta_B$ (that accounts for the extra $\frac{1}{4}$ factor), we obtain Bob's total sifted key rate:

\begin{equation}\label{qber3}
\begin{split}
&R_e^s = \frac{1}{4}  \sum_{j = H,V,D,A} \frac{1}{4}\sum_{\theta= 0^{\circ},45^{\circ},90^{\circ},135^{\circ}} R_e(j | \theta)
\end{split}
\end{equation}

The error rates conditioned on Alice sending $H$-polarized light and Eve applying $\theta_B \in \{0~\degree,45~\degree,90~\degree,135~\degree\}$ rotation can be calculated similar to \cref{error} in previous section. They are:
\begin{equation} \label{eq6}
\begin{split}
E_{H| 0^{\circ}} & \approx P_{c}^{e} \mathrm{P_{v}(H|0^{\circ})} +  P_{w}^{e}\mathrm{P_{v}(V|0^{\circ})} \\
& + P_{nc}^{e}[\mathrm{P_v(D|0^{\circ})} + \mathrm{P_v(A|0^{\circ})}]  \\
& + (1 - P_{c}^{e} - P_{w}^{e} - P_{nc}^{e})(c_v - \frac{c_v c_h}{2})
\end{split}
\end{equation}

\begin{equation} \label{eq7}
\begin{split}
E_{H| 90^{\circ}} & \approx P_{c}^{e} \mathrm{P_{h}(H|90^{\circ})} +  P_{w}^{e}\mathrm{P_{h}(V|90^{\circ})}\\
&+ P_{nc}^{e}[\mathrm{P_h(D|90^{\circ})}+ \mathrm{P_h(A|90^{\circ})}]  \\
& + (1 - P_{c}^{e} - P_{w}^{e} - P_{nc}^{e})(c_v - \frac{c_v c_h}{2})
\end{split}
\end{equation}

\begin{equation} \label{eq8}
\begin{split}
E_{H| 45^{\circ}} & \approx P_{c}^{e} \mathrm{P_{a}(H|45^{\circ})} +  P_{w}^{e}\mathrm{P_{a}(V|45^{\circ})}\\
&+ P_{nc}^{e}[\mathrm{P_a(D|45^{\circ})}+\mathrm{P_a(A|45^{\circ})}]  \\
&  + (1 - P_{c}^{e} - P_{w}^{e} - P_{nc}^{e})(c_v - \frac{c_v c_h}{2})
\end{split}
\end{equation}

\begin{equation} \label{eq9}
\begin{split}
E_{H| 135^{\circ}} & \approx P_{c}^{e} \mathrm{P_{d}(H|135^{\circ})} +  P_{w}^{e}\mathrm{P_{d}(V|135^{\circ})}\\
&+ P_{nc}^{e}[\mathrm{P_d(D|135^{\circ})}+ \mathrm{P_d(A|135^{\circ})}]  \\
& + (1 - P_{c}^{e} - P_{w}^{e} - P_{nc}^{e})(c_v - \frac{c_v c_h}{2})
\end{split}
\end{equation}

Here, $\mathrm{P_{i}(j|\theta_B)}$ is the probability that outcome $i$ is selected by Bob after squashing given that Eve has sent $j$-polarized light which was rotated by Bob by an angle $\theta_B$. So, modifying \cref{phvi} we get,
\begin{equation} \label{eq10}
\begin{split}
\mathrm{P_v(H|\theta)} & = [p_v(H|\theta) - \frac{p_h(H|\theta)p_v(H|\theta)}{2}]\\
& \times[1-p_d(H|\theta)][1-p_a(H|\theta)]
\end{split}
\end{equation}

The total QBER in Eve's presence becomes:

\begin{equation}
\begin{aligned}
QBER_e^s = \frac{1}{4R_e} \sum_{j=H,V,D,A} \frac{1}{4}\sum_{\theta= 0^{\circ},45^{\circ},90^{\circ},135^{\circ}} E_{j | \theta}
\end{aligned}
\end{equation}

In a plausible scenario, if Bob applies $\theta_B = 45~\degree$ and expects an incoming \emph{H} polarized light from Alice, he will be expecting a click in his \emph{d} detector. But if the light is coming from Eve, it will be directed towards, the \emph{H} attack angle where the \emph{h} detector has the highest efficiency. This increases the error rate. Simulations also verify that, scrambling the role of detectors can smoke out Eve's presence. 

%\section{Sifted key rate in Eve's Absence}
%\label{expected_skr}
%\ssc{if this equation is the same as my paper, lets just cite it somewhere and delete this section}\rf{this equation is not present in your paper, but this is not super important to show this work} Here, we will derive the expected sifted key rate when Eve is absent. Let $p_{corr} ~(p_{in})$ be the probability that Bob gets a detection in the correct (incorrect) detector given he measures in the correct basis. Let $p_{nc}$ be the probability where Bob gets a click on the non-compatible basis. Their expressions are similar to the cases in equation \ref{rcb}. Let $R_{ab}(j)$ be Bob's click rate when Alice sends $j$ polarized light. Thus, we find the expression for $R_{ab}(H)$ following the work in equation \ref{phvi}
%
%\begin{equation}\label{rab1}
%\begin{split}
%R_{ab}(H) \approx (1-p_{nc})(1-p_{nc})(p_{corr}+p_{in}-p_{corr}.p_{in})
%\end{split}
%\end{equation}
%
%Thus, the expression for expected sifted key rate in Eve's absence will be:
%\begin{equation}
%R_{ab} = \frac{1}{4} \sum_{j = H,V,D,A} R_{ab}(j)
%\label{rab_avg}
%\end{equation}
%%Steps for typesetting document with bibtex
%%1. Latex
%%2. Bibtex
%%3. Latex
%%4. Latex

\end{document}